\font\open=msbm10			%Get a font
\def\OR{\mbox{\open\char82}}		%Define an open R
\def\cosec{\mbox{cosec}}		%Define cosec

\documentstyle[preprint,pre,eqsecnum,aps]{revtex}
\input epsf
\begin{document}
\draft

\date{\today}
\title{Phase Diagrams for Deformable Toroidal and Spherical Surfaces with
Intrinsic Orientational Order}
\author{R.M.L.Evans}
\address{Theoretical Physics Group \\
Department of Physics and Astronomy \\
The University of Manchester, M13 9PL, UK}
\maketitle

\begin{abstract}
A theoretical study of toroidal membranes with various degrees of intrinsic
orientational order is presented at mean-field level. The study uses a simple
Ginzburg-Landau style free energy functional, which gives rise to a rich
variety of physics and reveals some unusual ordered states. The system is found
to exhibit many different phases with continuous and first order phase
transitions, and phenomena including spontaneous symmetry breaking, ground
states with nodes and the formation of vortex-antivortex quartets. Transitions
between toroidal phases with different configurations of the order parameter
and different aspect ratios are plotted as functions of the thermodynamic
parameters. Regions of the phase diagrams in which spherical vesicles form are
also shown.
\end{abstract}

\section{Introduction}

  The bilayer fluid membranes, which can form spontaneously when molecules with
hydrophobic and hydrophilic parts are introduced to water, are a popular topic
of research. In theoretical studies of such membranes, it is usual to work on
length scales much larger than the molecular size, so that they can be
considered as continuous surfaces; two-dimensional curved spaces embedded in
Euclidean three-space. Features commonly incorporated into mathematical
models of membranes include bending rigidity $\kappa$ (disaffinity for extrinsic
curvature), Gaussian curvature modulus $\kappa_{G}$ (disaffinity for intrinsic
curvature), spontaneous curvature due to membrane asymmetries, and for closed
surfaces, constraints of constant volume, constant area and constant difference
in area between the inner and outer layers. In recent years, interest has grown
in membranes whose molecules have orientational order within the surface. For
instance, a smectic-A liquid crystal membrane, whose molecules have aliphatic
tails which point in an average direction normal to the local tangent plane of
the membrane, can undergo a continuous phase transition to smectic-C phase, in
which the tails tilt. The tilt is described by a two-component vector order
parameter within the surface, given by the local thermal average of vectors
parallel to the tails, projected onto to local tangent plane. If such a
membrane has intrinsic (Gaussian) curvature, its orientational order is
frustrated, since a parallel vector field cannot exist in a curved space.
Ordering is further frustrated on a closed surface of spherical topology (genus
zero), as
an intrinsic vector field must have topological defects whose indices sum to
the Euler number of the surface; in this case, two. {\em ie.\ }there must be at
least two vortices in the order parameter field on a sphere. The order
parameter field vanishes at these defects to avoid infinite gradients.
The defects are energetically costly but topologically unavoidable.
The energetics of a smectic-C, genus zero membrane were studied by MacKintosh
and Lubensky in \cite{MacKintosh91}. There are other types of orientational
order. Nematic order is similar to vector order except that its order parameter
is locally invariant under rotations through $180^o$. Hence nematics can
form defects of index $\frac{1}{2}$. Hexatic order is locally invariant under
rotations through $60^o$, as this is the molecular `bond' angle. The
notion may be generalized to fluid phases with $n$-atic orientational order
being locally invariant under rotations through $\frac{2\pi}{n}$ and forming
defects of index $\frac{1}{n}$. If a vector (`monatic') is
represented by an arrow then we may picture an $n$-atic as an $n$-headed arrow
or an $n$-spoked wheel. For a theoretical study of $n$-atic fluid membranes of
genus zero, see \cite{Park92}.

  In this paper, theoretical findings on $n$-atic fluid membranes of toroidal
topology are presented. As the torus has genus one and Euler number zero, it is
possible to arrange an intrinsic vector field on a torus without
topological defects. ({\em ie.\ }furry tori can be brushed smooth.)
Therefore it appears plausible that $n$-atic membranes with large order
parameter coupling compared with bending rigidity may energetically favour a
toroidal topology to a spherical one. The object of embarking on this research
was to confirm or refute this hypothesis. As will be demonstrated, it turns out
to be true, although the arrangements of ordering in the toroidal surface are
far more diverse than expected.

  Throughout this paper, a number of approximations are used. One approximation
is the use of mean field theory. That is, no thermal fluctuations are
considered.
So the configuration adopted by the system is assumed to be that which
minimizes its free energy. Also, the lowest Landau level approximation is
employed in constructing the order parameter field, as discussed in section
\ref{Landau}.
The validity of these approximations is investigated
in section \ref{validity}. Both of these approximations give rise to apparent
phase transitions between separate, clearly defined states of the system, some
of which may in fact be blurred by thermal fluctuations and higher Landau
levels, to the point where
they disappear. It should therefore be understood that, where reference is made
to phase transitions, it is in the context of mean field theory and the lowest
Landau level approximation.
Before studying the thermal physics, it will be necessary to introduce some
tensor calculus. This is needed to describe the behaviour of an $n$-atic field
living within a two-dimensional curved space, as well as the intrinsic and
extrinsic elastic properties of the space ({\em ie.\ }the membrane) itself. In
the absence of intrinsic ordering ({\em ie.\ }under the influence of the
membrane's elasticity alone), one might expect the equilibrium shape of a
toroidal vesicle to be axisymmetric. However, such a vesicle is only neutrally
stable with respect to one particular mode of non-axisymmetric deformation.
This Goldstone mode is discussed in section \ref{Goldstone}, in which a further
approximation is introduced, limiting the range of vesicle shapes to be
explored. In the sections that follow, the ground-state configurations and
energies of the order parameter field are calculated for each of these shapes,
subject to the lowest Landau level approximation.
Although some of these shapes are never energetically favoured, the behaviour
of the field living on them is of interest, if only academic. Finally, in
section \ref{PD}, the total free energy (consisting of order-parameter field
and membrane elasticity parts) is minimized with respect to vesicle shape, to
find the equilibrium configuration of the system as a function of temperature
and the other thermodynamic parameters. The result is a set of phase diagrams
for various values of $n$, which contain phases of vesicles with different
shapes and ordering. The shapes should be experimentally observable, although
the intrinsic ordering is most likely not.

\section{Values of $n$}

  $n$ is the order of rotational symmetry. As stated earlier, the fluid membrane
is locally invariant under rotations through $\frac{2\pi}{n}$, and the $n$-atic
field may be represented by a set of $n$-spoked wheels tessellating. Clearly
$n$ is integer. Furthermore, it must be one of the set of numbers for which
regular $n$-gons tessellate. That is, $n\in \{ 1,2,3,4,6 \}$. These
are the values of $n$ which Park et.\ al.\ studied in \cite{Park92}. However,
consider a triangular ($n=3$) lattice with long-range {\em orientational}, but
no {\em positional} order; the $n=3$ analogue of a hexatic lattice. Bonds
between neighbouring molecules lie at angles
which are integer multiples of $\frac{\pi}{3}$. Its symmetry properties in the
absence of positional order are therefore indistinguishable from its dual
lattice, the hexatic, and it will form defects of index $\frac{1}{6}$, in
preference to the more energetically expensive index $\frac{1}{3}$ defects. To
convince oneself of this, it might be helpful to attempt to draw a lattice
which has $120^o$ rotational symmetry, but does not have $60^o$ rotational
symmetry. This is only possible if one keeps strict account of the positions of
the lattice sites and polygons ({\em ie.\ }in the absence of dislocations).
The following research is therefore conducted for $n\in \{ 1,2,4,6 \}$.

\section{Notation}

  The $2D$ curved surface is embedded in Euclidean $3$-space.
If {\boldmath $\sigma$}
is a two-dimensional coordinate of a point $P$ on the surface then
{\boldmath $R(\sigma)$}
is the position vector of $P$ in the embedding space
relative to some origin.
Coordinate-dependent unit basis vectors {\boldmath $e_{a}$}, where $a$ takes
the values $1$ and $2$ to label the coordinates, are defined in the directions
of $\partial_{a}\mbox{\boldmath $R$}$.
A coordinate-invariant metric tensor is defined on
the Riemannian $2$-space, with components
\mbox{$g_{ab} = \partial_{a} \mbox{\boldmath $R$} \cdot \partial_{b}
\mbox{\boldmath $R$}$}, from which the area of the closed surface
is calculated:
\begin{equation}
\label{area}
{\cal A} = \int d^2 \mbox{\boldmath $\sigma$}\sqrt{g}
\end{equation}
where $g \equiv \det{\underline{\underline{g}}}$. Indices are raised and
lowered by the metric and its inverse, $g^{ab}$, in the standard manner.
The extrinsic curvature tensor has components
$K_{ab} = \mbox{\boldmath $N$} \cdot
\partial_{a} \partial_{b} \mbox{\boldmath $R$}$,
where {\boldmath $N$} is a unit vector normal to the surface. (N.B. Dot
products are evaluated in $\OR ^3$.)
Its trace, $K^a_{a}$, is the `total curvature'. Readers may be familiar with the
{\em mean} curvature which is half of this quantity. The intrinsic or Gaussian
curvature is given by $K = \det{\underline{\underline{K}}}$.

  As an example, the above quantities are calculated for a sphere of radius
$R$, using spherical polar coordinates $(\theta,\phi)$. The metric tensor may be
represented by a matrix:
\begin{displaymath}
  \underline{\underline{g}} \doteq
  \left( \begin{array}{cc}
           g_{11} & g_{12} \\
           g_{21} & g_{22}
         \end{array}  \right) =
  R^2 \left( \begin{array}{cc}
           1 & 0 \\
           0 & \mbox{sin}^2 \theta
         \end{array}  \right) ,
\end{displaymath}
with determinant $g=R^4 \mbox{sin}^2 \theta$. Hence the area element for
integration is $\sqrt{g}\, d^2 \mbox{\boldmath $\sigma$} = \sin{\theta} \,
d\theta \, d\phi $, which is just the Jacobian for spherical polars. Any
continuous, doubly differentiable
surface has two principal radii of curvature, $R_{1}$ and $R_{2}$ at each
point. These are the radii of curvature of the geodesics passing through that
point which have respectively the greatest and least curvature as measured in
the Euclidean embedding space. It can be shown \cite{Rutherford}
that these two geodesics are
always perpendicular. If the particular coordinate basis used corresponds to
these two perpendicular directions, then the curvature tensor will be
diagonal, and given by
\begin{displaymath}
  \underline{\underline{K}} \doteq
  \left( \begin{array}{cc}
           K_{1}^1 & K_{1}^2 \\
           K_{2}^1 & K_{2}^2
         \end{array}  \right) =
  \left( \begin{array}{cc}
           \frac{1}{R_{1}} & 0 \\
           0 & \frac{1}{R_{2}}
         \end{array}  \right) .
\end{displaymath}
Of course, on a sphere, curvature is constant.
Therefore $K_{a}^{b}=-\frac{\delta_{a}^{b}}{R}$ where the sign convention
denotes the direction of the normal. However, it should be noted that, without
one index raised, the components are not constant:
\begin{displaymath}
  \left( \begin{array}{cc}
           K_{11} & K_{12} \\
           K_{21} & K_{22}
         \end{array}  \right) =
  -R \left( \begin{array}{cc}
           1 & 0 \\
           0 & \mbox{sin}^2 \theta
         \end{array}  \right) .
\end{displaymath}
Clearly, on a sphere, mean curvature is $-\frac{1}{R}$ and Gaussian curvature is
$\frac{1}{R^2}$.

  Recall that the intrinsic $n$-atic order in the system is orientational, not
positional, and that it is locally invariant under rotations through
$\frac{2\pi}{n}$. As this orientational order is intrinsic to a 2-dimensional
space, in can be represented by a complex order parameter in the following way.
Let $\Theta$ be the local angle between {\boldmath $e_{1}$} and the orientation
of ordering. For vector order, this orientation is easily defined. For
instance, in a smectic-C membrane, it is the direction of a molecule's
aliphatic tail projected onto the local tangent plane of the membrane. However,
in a nematic liquid crystal, $\Theta$ may take one of two values for each
molecule, depending on which end of the molecule one chooses to measure from.
For our purposes, it does not matter which of these two values is assigned to
$\Theta$ at each point. For bond-angle orientational order, {\em eg.\ }hexatic,
there is an $n$-fold ambiguity in the definition of $\Theta$ but again, it
does not matter which of the $n$ orientations is chosen. It is
conventional in this case to define $\Theta$ at a given molecule as the angle
between the basis vector {\boldmath $e_{1}$} and the line joining that molecule
to its nearest neighbour. Having defined $\Theta$, the order
parameter is given by
\begin{equation}
\label{psi}
  \psi(\mbox{\boldmath $\sigma$}) = 
  \left< \exp ( i n \Theta(\mbox{\boldmath $\sigma$}) ) \right>
\end{equation}
where $<>$ denotes a thermal average. Note that $\psi$
has the appropriate rotational invariance, since a rotation of the
orientation of ordering through an angle $\chi$ within the surface is
represented by a rotation of $\psi$ through $n\chi$ in the Argand plane. Hence
rotating the orientation of ordering through $\frac{2\pi}{n}$ corresponds
an identity transformation in the complex plane.

\section{The Model}

  The model used is identical to that in \cite{Park92}. There is zero
spontaneous curvature as the bilayers considered are symmetric, and no
constraint on bilayer area difference.
The vesicle's volume is also unconstrained
since small solvent molecules may diffuse freely in and out,
so long as the solvent is very pure; containing no free ions, which
would lead to osmotic pressure. The membrane's area ${\cal A}$ is fixed by
the number of constituent molecules. The free energy functional
\begin{equation}
\label{F}
  F[\psi(\mbox{\boldmath $\sigma$})
  ,\mbox{\boldmath $R$}(\mbox{\boldmath $\sigma$})]/T = \int d^2
  \mbox{\boldmath $\sigma$}\sqrt{g}\,
  \left( \,r|\psi |^2 + \mbox{$\frac{1}{2}$}u|\psi |^4
  +CD^{a}\psi D^{*}_{a}\psi ^*
  +\mbox{$\frac{1}{2}$}\kappa (K^{a}_{a})^2 + \kappa_{G} K \,\right)
\end{equation}
closely resembles the Ginzburg-Landau free energy for classical
superconductors. For a discussion of how the
thermodynamic coefficients relate to real
physical and chemical quantities, see for instance \cite{David}.
As defined in Eq.(\ref{psi}), the phase of $\psi$ depends on the
choice of basis vectors. Hence the derivatives $D_{a}$ in Eq.(\ref{F}) must be
covariant in order for the quantity $D^a\psi D^*_{a}\psi^*$ to be physical.
Let the basis vectors be rotated
through an angle $-\chi(\mbox{\boldmath $\sigma$})$. Consequently,
\begin{displaymath}
\begin{array}{lclcl}
  \psi &\rightarrow& \psi' &=& \psi e^{in\chi(\mbox{\boldmath $\sigma$})}\\
  \partial_{a} &\rightarrow& \partial'_{a} &=&
    \Lambda_{a}^{\:c}\partial_{c}\\
  g_{ab} &\rightarrow& g'_{ab} &=& \Lambda_{a}^{\:c}\Lambda_{b}^{\:d}
    g_{cd}
\end{array}
\end{displaymath}
where $\Lambda_{a}^{\:c}(\mbox{\boldmath $\sigma$})$ are the components of
the rotation matrix. Given that $D_{a}=\partial_{a}-inA_{a}$, demanding that
$D^a\psi D^*_{a}\psi^*$ is invariant
under the gauge transformation implies that the components of {\boldmath $A$}
transform like
\begin{equation}
\label{transform}
\begin{array}{lclcl}
  A_{a} &\rightarrow& A'_{a} &=& \Lambda_{a}^{\:c}
   (A_{c}+\partial_{c}\alpha)\,.
\end{array}
\end{equation}
So {\boldmath $A$} undergoes the usual transformation, $\partial_{a}\alpha$, of
a vector potential in a gauge-invariant derivative. But, in addition there is a
coordinate rotation, $\Lambda_{a}^{\:c}$ since, in this complex number
representation, a rotation of the Argand plane is always accompanied by a
rotation of the basis vectors. The `spin connection', $A_{a}=\mbox{\boldmath
$e$}_{1} \cdot \partial_{a} \mbox{\boldmath $e$}_{2}$ has the appropriate
transformation \mbox{property Eq.(\ref{transform})}. On the sphere,
$A_{\theta}=0$ and $A_{\phi}=-\cos{\theta}$.

\section{The Curvature Energy Terms}
\label{Goldstone}

  Above the continuous transition to the disordered phase, $\psi$ vanishes and
only the last two terms in the free energy Eq.(\ref{F}) survive. These
constitute the Helfrich Hamiltonian, describing the elastic energy of a
constant area membrane.

  The {\em intrinsic} curvature energy, being proportional to the integral of
Gaussian curvature, is a {\em topological} invariant, according to the
Gauss-Bonnet formula
\begin{equation}
\label{GaussBonnet}
  \int K \, d{\cal A} = 2\pi\chi
\end{equation}
where $\chi$ is the Euler number of the closed surface \cite{Weeks}. For an
orientable
surface, $\chi=2(1-G)$, where $G$ is the surface's genus; {\em ie.\ }how
many `handles' it has. Hence the intrinsic curvature energy is zero for a torus
and $4\pi\kappa_G$ for a sphere.

\begin{figure}
  \epsfxsize=15cm
  \begin{center}
  \leavevmode\epsfbox{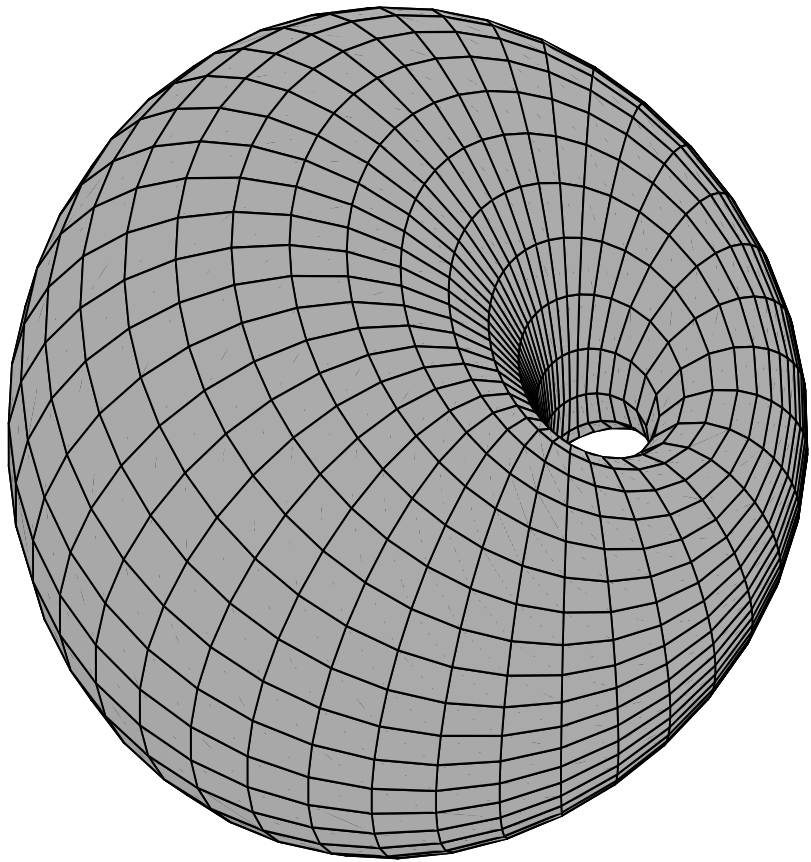}
  \caption{The surface produced by a conformal transformation of an
  axisymmetric, circular cross-section torus.}
  \label{piccy}
  \end{center}
\end{figure}

  The {\em total} curvature energy,
$\frac{\kappa}{2}\int(K^a_{a})^2\,d{\cal A}$, is
a {\em conformal} invariant \cite{Fourcade92}. {\em ie.\ }it is invariant under
shape changes of the form
\begin{equation}
\label{conformal}
  \mbox{\boldmath $R$} \rightarrow \mbox{\boldmath $R'$} = 
  \frac{\frac{\mbox{\boldmath $R$}}{|R|^2}+\mbox{\boldmath $\beta$}}{\left|
  \frac{\mbox{\boldmath $R$}}{|R|^2}+\mbox{\boldmath $\beta$}\right| ^2}
\end{equation}
where {\boldmath $\beta$} is a constant vector which parameterizes the conformal
transformation \cite{Julicher93}. A sphere has total curvature energy
$8\pi\kappa$ and
its shape is not changed by such a transformation.
This is not true for tori. An axisymmetric, circular cross section torus may be
deformed
by a conformal transformation. As a consequence of the circular cross-section,
a component of {\boldmath $\beta$} parallel to the
symmetry axis serves only to change the torus's size. Hence there remains only
a one-parameter family of non-trivial conformal transformations. With
{\boldmath $\beta$} lying in the symmetry plane, the surface is deformed
into a non-axisymmetric torus such as the one in
Fig.(\ref{piccy}), and as {\boldmath $\beta$} increases in magnitude, the
asymmetry
increases until, in the limit, the torus becomes a perfect sphere with an
infinitesimal handle.
The handle, although infinitesimal, must contain an
intrinsic curvature energy of $-4\pi\kappa_G$ and finite total curvature
energy. A circular, axisymmetric torus of aspect ratio (ratio of radii of
generating \mbox{circles) $\alpha$ }\nopagebreak has total curvature energy
\nopagebreak
\begin{equation}
\label{curvenergy}
  \frac{\kappa}{2}\int(K^a_{a})^2\,d{\cal A} =
  \frac{2\pi^2\alpha^2\kappa}{\sqrt{\alpha^2-1}},
\end{equation}
which is a minimum at
$\alpha=\sqrt{2}$ , the Clifford torus, and infinite at $\alpha=1$, where the
torus becomes self-intersecting. Willmore conjectured \cite{Willmore}
that Clifford's torus and its conformal transformations have the
{\em absolute} minimum curvature energy of any surface of toroidal topology; a
conjecture
that is difficult to prove, as the most general toroidal surface cannot easily
be parameterized: tori can even be knotted. Accordingly, this paper will not
explore the whole of phase space. The discussion will be restricted to
axisymmetric, circular cross-section tori and their conformal transformations,
with the assumption that these are close to the ground state shapes. To be
consistent, the ground-state genus-zero shape is approximated
by a sphere: the deviations from sphericity calculated in \cite{Park92} are not
included.

\section{The Lowest Landau Level Approximation}
\label{Landau}

  The field gradient energy term in Eq.(\ref{F}) may be re-formulated. Using
integration by parts, it is simple to show that, on a closed surface,
\begin{equation}
\label{gradient}
  \int d^2 \mbox{\boldmath $\sigma$} \sqrt{g}\, g^{ab} D_{a}\psi D^*_{b}\psi^*
  = -\int d^2 \mbox{\boldmath $\sigma$} \psi^* D_{a}(\sqrt{g}\,g^{ab}D_{b}\psi).
\end{equation}
Let $\phi_{p}$ be eigenfunctions satisfying the Hermitian equation
\begin{equation}
\label{eigen}
  -\frac{D_{a}(\sqrt{g}\,D^a)\phi_{p}}{\sqrt{g}} = \Lambda \phi_{p}
\end{equation}
and the orthonormalization condition
\begin{equation}
\label{normalization}
  \int d^2\mbox{\boldmath $\sigma$} \sqrt{g}\, \phi^*_{p}\phi_{q} = 
  \frac{\cal A}{4\pi}\delta_{pq} .
\end{equation}
These are the Landau levels. $\psi$ may be expanded in this complete set of
functions. However, close to the mean-field transition, it is a good
approximation to use only the Landau levels with the lowest eigenvalue,
$\Lambda_{0}$. As is
common in such problems, there is a degenerate set of lowest Landau levels.

  $\Lambda$ scales inversely with area. So let us define a scale-invariant
eigenvalue $\lambda$, by
\begin{equation}
\label{lambda}
  \Lambda \equiv \frac{4\pi}{\cal A} \lambda .
\end{equation}

  The free energy becomes
\begin{displaymath}
  F/T = \int \left\{ \left( r+\frac{4\pi C \lambda_{0}}{\cal A} \right)
  |\psi|^2 + \frac{u}{2}|\psi|^4 \right\} d{\cal A} + \mbox{curvature energy}
\end{displaymath}
where
\begin{equation}
\label{expansion}
  \psi = \sum_{p} a_{p} \phi_{p}
\end{equation}
and $a_{p}$ are complex constants.
Note that the mean field transition has been lowered from $r=0$ to $r=r_{c}$
where $r_{c}\equiv-\frac{4\pi C \lambda_{0}}{\cal A}$. On an intrinsically flat
surface,
$\lambda_{0}=0$ but it is finite and positive-definite on a surface with finite
Gaussian curvature. So ordering is frustrated by intrinsic curvature.

\section{Solution of the Eigenfunction Equation}

  Eq.(\ref{eigen}) may be re-written as the differential equation
\begin{equation}
\label{eigen2}
  \left[g^{ab}\right] \partial_{a} \partial_{b} \phi_{p}
  + \left[(\frac{\partial_{b}-2inA_{b}}{\sqrt{g}})(\sqrt{g}\,g^{ab})\right]
  \partial_{a} \phi_{p}
  + \left[ \frac{-inD_{a}}{\sqrt{g}}(\sqrt{g}\,g^{ab}A_{b}) \right] \psi_{p}
  = -\frac{4\pi}{\cal A} \lambda \phi_{p} .
\end{equation}
This is still general to any surface.

\subsection{Sphere}
\label{sphere}

  On a sphere, in spherical polar coordinates $(\theta,\phi)$,
Eq.(\ref{eigen2}) becomes
\begin{equation}
\label{eigensphere}
  \left(\frac{\partial^2}{\partial\theta^2} + \cosec^2\theta
  \frac{\partial^2}{\partial\phi^2}+\cot\theta\frac{\partial}{\partial
  \theta} + 2in\cosec\theta\cot\theta\frac{\partial}{\partial\phi}
  - n^2 \cot^2 \theta \right)\phi_{p} = -\frac{4\pi\lambda}{\cal A}\phi_{p}
\end{equation}
which is solved by $\lambda_{0}=n$ and $\phi_{p}$ of the form
\begin{equation}
\label{sphereeigenfuncs}
  \phi_{p} \propto \sin^{n+j} \left(\mbox{$\frac{\theta}{2}$}\right)
  \cos^{n-j} \left(\mbox{$\frac{\theta}{2}$}\right) e^{ij\phi}
\end{equation}
where $p$ takes integer
values from $-n$ to $n$. These functions form an alternative basis set to that
used in \cite{Park92} and are linear combinations of that set. They are given
in \cite{ONeill} in a different gauge. The functions have $2n$ zeros, which are
topological defects of index $\frac{1}{n}$. The $(2n+1)$-fold degeneracy
reflects the freedom of the defects to lie anywhere on the surface at no energy
cost to order $\psi^2$.

\subsection{Axisymmetric Torus}
\label{axisym}

\begin{figure}
  \epsfxsize=15cm
  \begin{center}
  \leavevmode\epsfbox{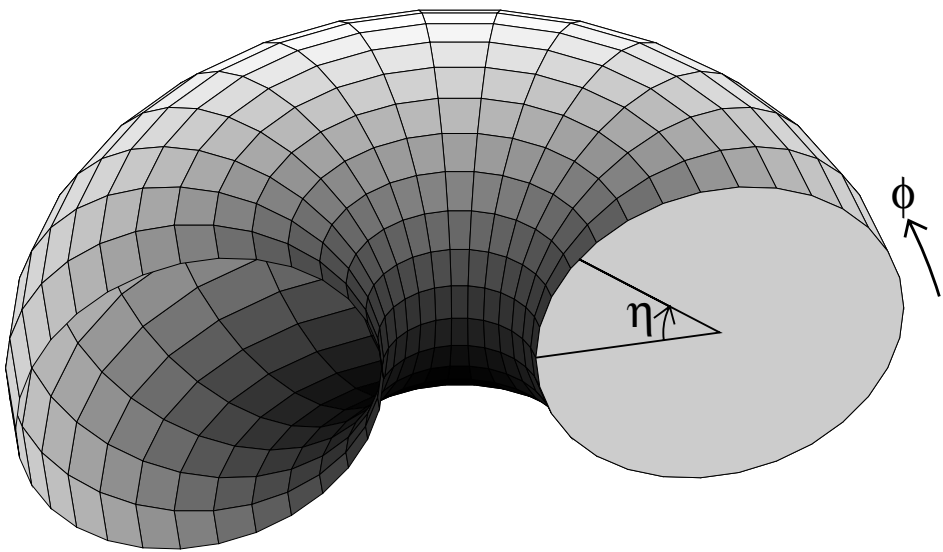}
  \caption{The angular coordinates $(\eta,\phi)$ defined on an axisymmetric,
  circular cross-section torus. For clarity, only half of the torus is shown.}
  \label{coords}
  \end{center}
\end{figure}

  On tori, $\phi_{p}$ and $\lambda$ depend on $\alpha$ and $\beta$.
Consider first the axisymmetric case ($\beta=0$). Let us define coordinates
$(\eta,\phi)$. The azimuthal angle is $\phi$, varying around the large
generating circle and subtended at the centre. The angle $\eta$ varies
around the smaller generating circle and is measured from the symmetry plane, as
shown in Fig.(\ref{coords}).
Let the small generating circle have radius $R_{0}$. In this coordinate system,
the following quantities are found:
\begin{eqnarray*}
  g_{ab} &\doteq& R_{0}^2 \left( \begin{array}{cc}
                                    1 & 0 \\
                                    0 & (\alpha-\cos\eta)^2
                                 \end{array} \right)          \\
  K_{a}^b &\doteq& \frac{1}{R_{0}} \left( \begin{array}{cc}
                                             -1 & 0 \\
                                       0 & \frac{\cos\eta}{\alpha-\cos\eta}
                                          \end{array} \right) \\
  A_{a} &\doteq& \left( \begin{array}{c}
                           0 \\
                           -\sin\eta
                        \end{array} \right)
\end{eqnarray*}
The Landau level equation Eq.(\ref{eigen2}) becomes
\begin{equation}
\label{eigentorus}
  \left[ \partial_{\eta}\partial_{\eta} + \frac{\partial_{\phi}\partial_{\phi}}
  {(\alpha-\cos\eta)^2} + \frac{\sin\eta}{\alpha-\cos\eta}\partial_{\eta}
  + \frac{2in\sin\eta}{(\alpha-\cos\eta)^2}\partial_{\phi}
  - n^2 \left(\frac{\sin\eta}{\alpha-\cos\eta}\right)^2 \right] \phi_{p}
  = \frac{4\pi\lambda}{\cal A}\phi_{p} ,
\end{equation}
which is manifestly invariant under $\eta\rightarrow-\eta$ with complex
conjugation, reflecting the
up-down mirror symmetry of the torus.
Note that changing variable from $\eta$ to $\theta\equiv(\eta-\frac{\pi}{2})$
and setting $\alpha$ to zero re-creates the sphere
equation (Eq.(\ref{eigensphere})).

  Eq.(\ref{eigentorus}) has separable variables and is solved by functions of
the form
\begin{equation}
\label{form}
  \phi_{p}(\eta,\phi)=f_{p}(\eta) e^{ip\phi}
\end{equation}
for integer $p$, where
\begin{equation}
\label{eigentorus2}
  \left[\frac{d^2}{d\eta^2}+\frac{\sin\eta}{(\alpha-\cos\eta)}\frac{d}{d\eta}
  - \left(\frac{p+n\sin\eta}{\alpha-\cos\eta}\right)^2\right]f_{p}
  = \frac{4\pi\lambda}{\cal A}f_{p}
\end{equation}
and $f_{p}$ is real.
It is immediately apparent from the form of Eq.(\ref{eigentorus2}) that
\begin{equation}
\label{symmetry}
  f_{p}(\eta) = f_{-p}(-\eta)
\end{equation}
and consequently $\phi_{p}$ and $\phi_{-p}$ are degenerate pairs of states.

  The ground-state solutions of Eq.(\ref{eigentorus2}) were found numerically
for various values of $p$ and $\alpha$, using Rayleigh-Ritz variational method.
Trial functions were synthesized from a truncated
Fourier series,
\begin{displaymath}
  f_{p}^{\mbox{\tiny trial}} = \sum_{\nu=0}^{\nu_{\mbox{\tiny max}}} \left(
  a_{\nu}\cos{\nu\eta} + b_{\nu}\sin{\nu\eta} \right) ,
\end{displaymath}
which has the appropriate periodic boundary conditions.
The Fourier coefficients were used as variational parameters to minimize the
integral
\begin{equation}
\label{variationalfunctional}
  \int_{0}^{2\pi} (\alpha-\cos\eta) f_{p}^{\mbox{\tiny trial}}
  \left[\frac{d^2}{d\eta^2}+\frac{\sin\eta}{(\alpha-\cos\eta)}\frac{d}{d\eta}
  - \left(\frac{p+n\sin\eta}{\alpha-\cos\eta}\right)^2\right]
  f_{p}^{\mbox{\tiny trial}}\:d\eta
\end{equation}
subject to Eq.(\ref{normalization}). The minimal value of
Eq.(\ref{variationalfunctional}) is an upper bound on $\lambda_{0}$ , and the
corresponding function $f_{p}^{\mbox{\tiny trial}}$ is approximately $f_{p}$ .
Using a Fourier series truncated after frequency $\nu_{\mbox{\tiny max}}$
creates a
$(2\nu_{\mbox{\tiny max}}+1)$-dimensional parameter space to explore in the
numerical
minimization. So CPU time increases very rapidly with $\nu_{\mbox{\tiny max}}$.
A second numerical method was also used, which involved
numerically `integrating'
Eq.(\ref{eigentorus2}) forward in $\eta$ from $\eta_{0}$ to $\eta_{0}+2\pi$ and
minimizing the resultant discontinuities in $f_{p}$ and its derivative, with
respect to the initial conditions and $\lambda$. This method is more accurate
than the Rayleigh-Ritz method but more difficult to use since, depending on the
starting conditions, it may find a non-ground-state solution. However, it was
useful in establishing the accuracy of the upper bounds. Using
$\nu_{\mbox{\tiny max}}=7$, $\lambda_{0}$ was found to an accuracy of 0.05\% 
in the worst case.

\begin{figure}
  \epsfxsize=15cm
  \begin{center}
  \leavevmode\epsfbox{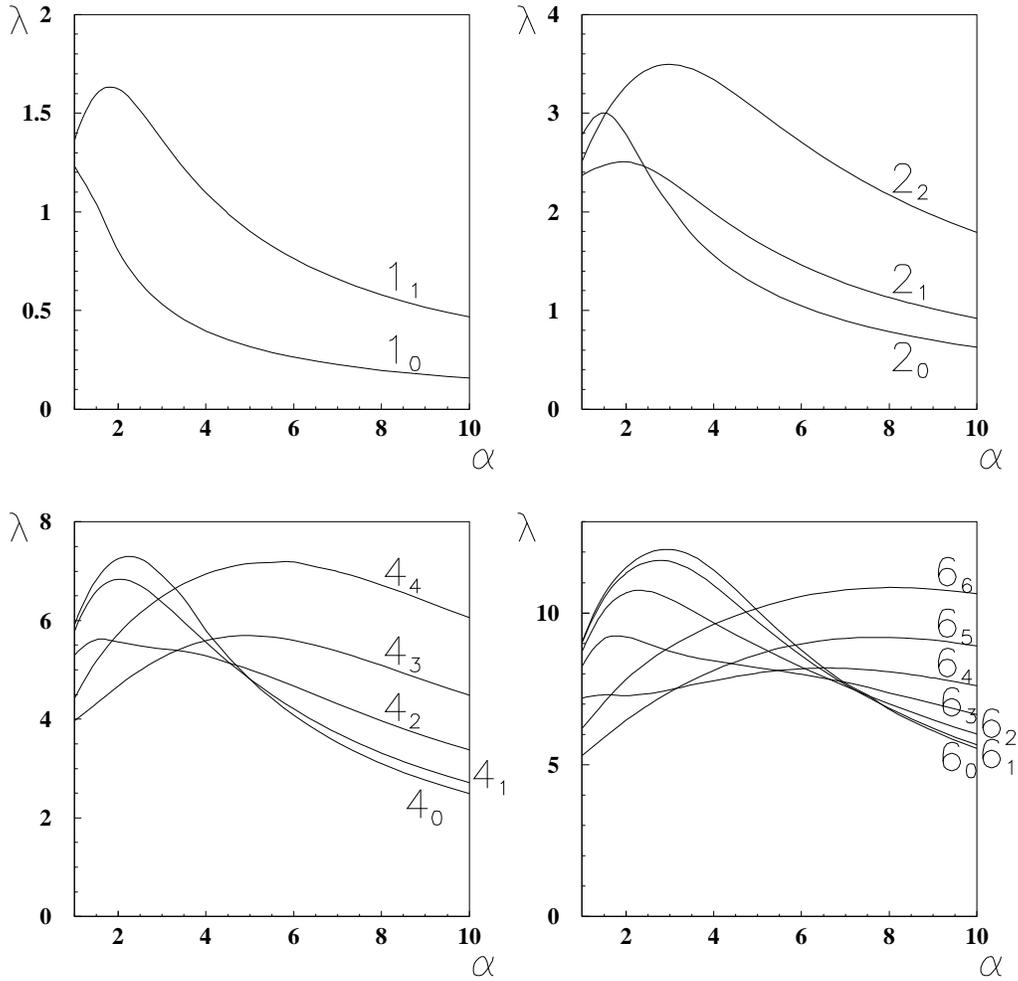}
  \caption{Graphs of $\lambda$ against $\alpha$ for axisymmetric, circular
  cross-section tori. The states are labelled $n_{p}$.}
  \label{lambdagraphs}
  \end{center}
\end{figure}

  In Fig.(\ref{lambdagraphs}), $\lambda(\alpha)$ is plotted for
$0\leq p\leq n$,
with data points at increments of 0.5 in $\alpha$ and cubic spline
interpolation. The states are labelled $n_{p}$. One might have guessed that
$n_{0}$ would be the ground state in each case. And indeed, for large aspect
ratios, when the torus is almost cylinder-like,
the $p=0$ states do have the lowest energy. These states are even
functions of $\eta$, being larger at $\eta=\pi$ than at $\eta=0$ where there is
more absolute intrinsic curvature,
and do not vary with $\phi$. {\em ie.\ }The field does not
twist with respect to the coordinate lines as $\phi$ increases. However, for
smaller $\alpha$, states with non-zero $p$ are energetically favoured.

\begin{figure}
  \epsfxsize=15cm
  \begin{center}
  \leavevmode\epsfbox{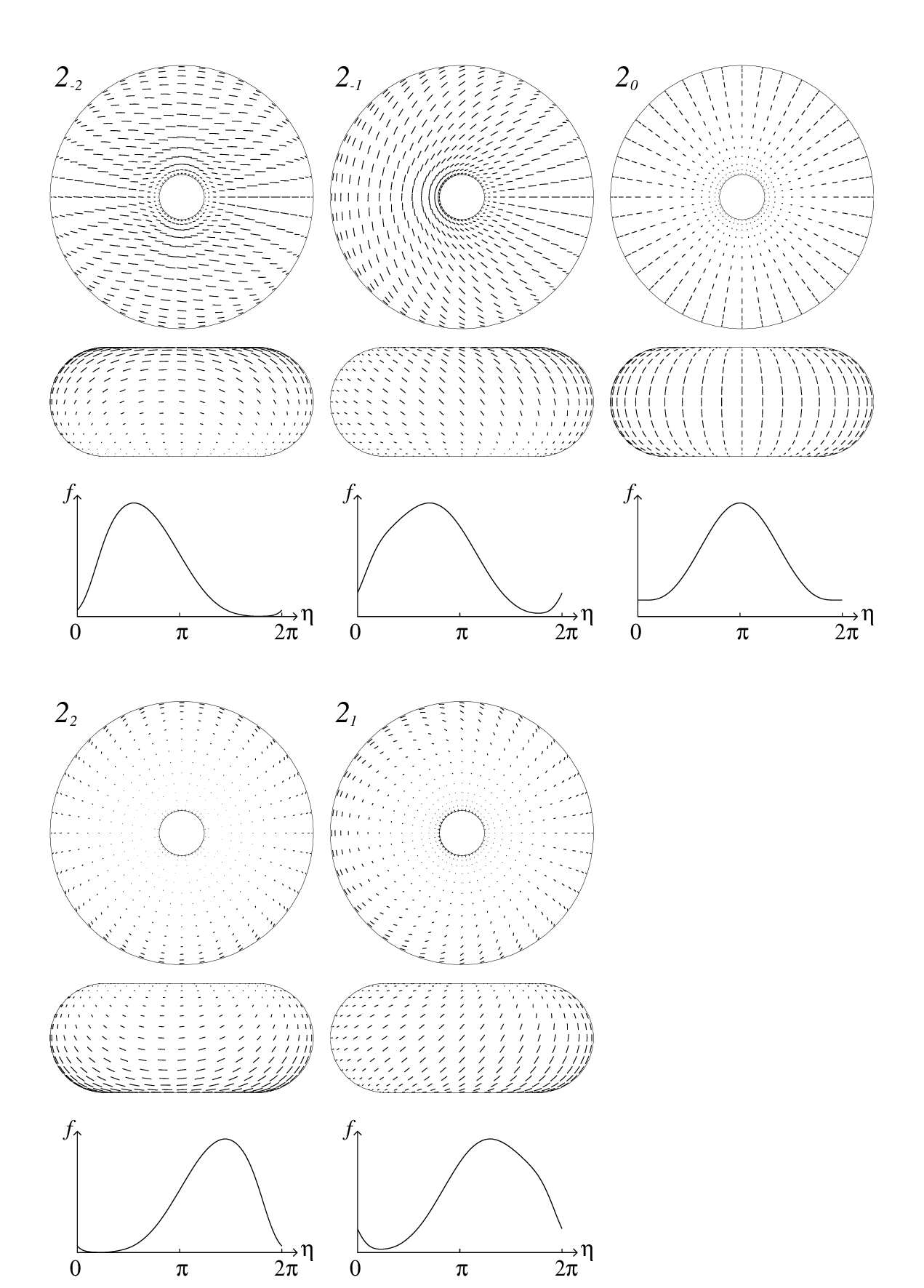}
  \caption{Various Landau levels of the $n=2$ (nematic) order-parameter field
  on a Clifford torus. States are labelled $n_{p}$ and shown in plan view and
  side view and $\eta$ dependence is given graphically.}
  \label{states}
  \end{center}
\end{figure}

  Before explaining the reason for this, it is necessary to describe these
states. Fig.(\ref{states}) depicts a selection of states for $n=2$ on a
Clifford torus. The nematic-order field is represented by sets of double-ended
lines directed along the orientations of order in the surface of the torus. The
length of each line is proportional to $|\psi|$ at that point. Note however
that the three-dimensional views of the toroidal surface cause the lines to be
foreshortened at oblique angles to the line of sight.
Two views are shown of each field on the toroidal surface, as well as a
graph of $f(\eta)$. Note that positive $p$ states have a large field on the
bottom of the torus and small on the top, and vice versa for negative $p$. This
asymmetry becomes more pronounced with increasing $p$, with increasing $n$, and
with decreasing $\alpha$. As stated earlier, ordering is frustrated by
intrinsic curvature. And the net amount of intrinsic curvature on the torus is
constant. But as $\alpha$ decreases, the curvature is redistributed. It becomes
more negative around the hole and more positive around the outside of the
torus, with the top and bottom remaining fairly flat. So this is where the
field prefers to reside and be parallel. Although in Eq.(\ref{form})
non-zero $p$ states appear to
twist with respect to the coordinate basis as $\phi$ increases, the basis
itself twists on the top
and bottom of the torus as the hole is circled. The two twists cancel. Note
that the diagram of the $2_{-2}$ state depicts a field which is as parallel as
possible on the top `face' of the torus. Of course, the field on the inner and
outer `rims' still contributes to the free energy so, as $\alpha$ increases,
the ground-state value of $p$ decreases by a series of `transitions'. This
is clear in Fig.(\ref{lambdagraphs}). These
should not be regarded as true phase transitions, since the lowest Landau level
approximation breaks down completely at these points. As a crossover between
eigenvalues is approached, the true state is a mixture of the two states.

  It should be emphasized that the tori studied here are true, curved tori,
embedded in 3-space. They should not be confused with a flat torus, which is a
plane with opposite edges identified, for which the above energetics would not
apply. The two have the same topology, but different local geometries.

\subsection{Non-Axisymmetric Torus}
\label{non-axisym}

  When $\beta\neq 0$, the Landau level equation (Eq.(\ref{eigen2})) becomes much
more complicated and its variables are no longer separable. However,
Rayleigh-Ritz variational method is again possible, as
\begin{equation}
\label{thing}
  \lambda_{0}(\alpha,\beta) \leq \int_{{\cal A}=4\pi}
  \phi^*_{\mbox{\tiny trial}} D_{a}(\sqrt{g}\,D^a ) \phi_{\mbox{\tiny trial}}
  \, d^2 \mbox{\boldmath $\sigma$} ,
\end{equation}
but the trial wavefunctions are now complex, periodic functions of two
variables. Thus, synthesizing them with a truncated, two-dimensional, complex
Fourier series up
to frequency $\nu_{\mbox{\tiny max}}$ produces an $(8\nu_{\mbox{\tiny max}}^2
+ 8\nu_{\mbox{\tiny max}} + 1)$-dimensional space of variational parameters to
be
numerically explored. Hence ground-state functions of the coordinates with high
frequency components cannot viably be found. This is unfortunate since
conformally transforming an axisymmetric torus with the coordinate system
described above, leaves the coordinates very unevenly distributed over the
surface, with most of the interval from $0$ to $2\pi$ in $\eta$ and $\phi$
covering only the small `handle' part of the torus and only a small range
covering most of the area. This problem is partly solved by a coordinate
transformation $(\eta,\phi)\rightarrow(\rho,\sigma)$ which preserves the
metric's diagonalization, where \nopagebreak
\begin{eqnarray*}
  [1-\beta(\alpha-1)]\tan\frac{\rho}{2} &=&
  [1-\beta(\alpha+1)]\tan\frac{\eta}{2} \\
  \left[1+\beta(\alpha+1)\right]\tan\frac{\sigma}{2} &=&
  [1-\beta(\alpha+1)]\tan\frac{\phi}{2}
\end{eqnarray*}
although this still leaves some high frequency components in the distribution
of the coordinates on the handle. Using these coordinates, Eq.(\ref{thing}) is
very complicated but two special cases are already known:
\begin{itemize}
  \item When $\beta=0$, the torus
is axisymmetric and $\lambda_{0}$ was found by the method in section
\ref{axisym}.
  \item When
$\beta=\frac{1}{R_{0}(\alpha+1)}$ , where $R_{0}$ is the radius of the small
generating circle, the torus becomes a sphere with an infinitesimal handle.
Hence $\lambda_{0}=n$ and the ground-state function is that of a sphere with
all $2n$ zeros at the handle.
\end{itemize}
Unfortunately, the problem is too computer-intensive to produce any useful
results for intermediate values of $\beta$. Using $\nu_{\mbox{\tiny max}}=3$ and
performing the minimization in
a 98-dimensional parameter space took a considerable amount of CPU time per
data point. At this level, no states of intermediate $\beta$ were found to have
a lower eigenvalue than the known axisymmetric or spherical states. Some
{\em local}
minima were found but these are very likely to be artifacts of the varying
reliability of the upper bound, which was high by around 10\% for $\beta=0$ and
60\% in the spherical limit. More research on this problem is planned. But for
the moment, it is necessary to work with the ansatz that:
\begin{itemize}
  \item  A spherical vesicle exists when its total free energy is lower than
  that of an axisymmetric torus.
  \item  An axisymmetric torus exists when its total free energy is lower than
  that of a sphere {\em and} its field energy alone is lower than that of a
  sphere.
  \item  A non-axisymmetric torus exists otherwise.
\end{itemize}

\section{The $\psi^4$ term}

\subsection{Axisymmetric Torus}
\label{axisym2}

  It was noted earlier that $\phi_{p}$ and $\phi_{-p}$ are degenerate to
quadratic order in $\psi$. So, from Eq.(\ref{expansion}), $\psi$ is a linear
combination of $\phi_{p}$ and $\phi_{-p}$,
where $p$ has the ground-state value for the given aspect ratio. The relative
proportions of
the two states are determined by the $\psi^4$ term in Eq.(\ref{F}). Let
\begin{displaymath}
  \Psi = a_{+}\phi_{p} + a_{-}\phi_{-p}
\end{displaymath}
be the function which minimizes $\int|\Psi|^4 d{\cal A}$ on a surface of area
$4\pi$ subject to $|a_{+}|^2+|a_{-}|^2=1$, and let J be the minimal value of
that integral. Then Eq.(\ref{F}) is minimized by $\psi=a\Psi$ in the regime of the
lowest Landau level approximation, where $a$ is real. The gauge invariance and
axial symmetry of the problem allow us to choose the phases of both $a_{+}$ and
$a_{-}$. Hence $\Psi$ is determined by a minimization with respect to just one
parameter.

\begin{figure}
  \epsfxsize=10cm
  \begin{center}
  \leavevmode\epsfbox{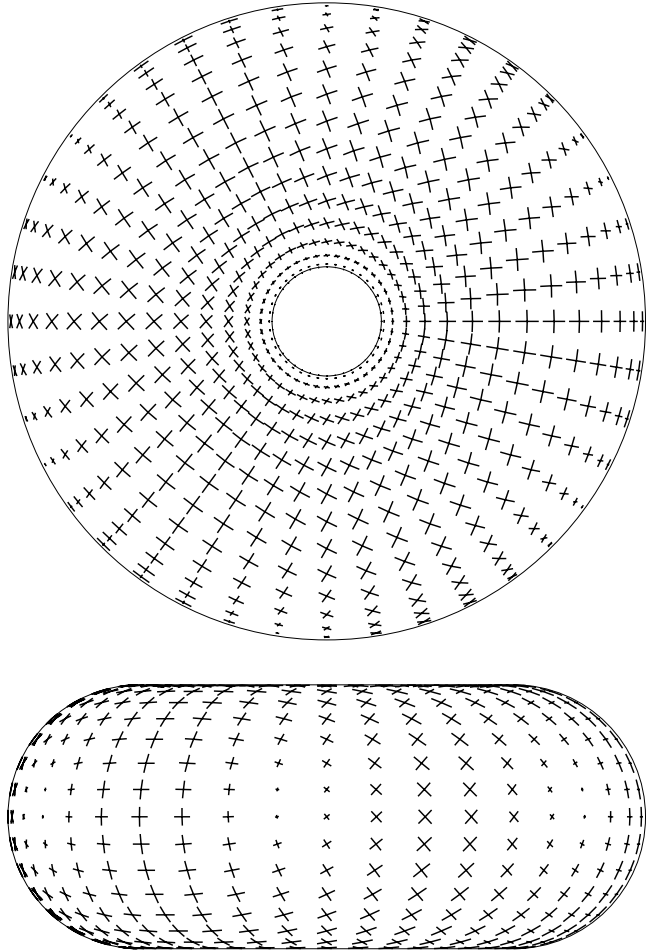}
  \caption{`Tetratic' ($n=4$) order-parameter field in the $4_{3}^S$
  configuration. The field is represented by crosses of four lines, directed
  along the orientations of order in the surface of the Clifford torus. The
  length of each line is proportional to $|\psi|$ at that point although the
  three-dimensional view of the toroidal surface causes the lines to be
  foreshortened at oblique angles to the line of sight. Notice that 3 of the 6
  defects of index $\frac{1}{4}$ are visible in the side view (where
  $|\psi|\rightarrow 0$), and that the field rotates through
  $\frac{n-p}{n}=\frac{1}{4}$ of a turn around the hole.}
  \label{n4p3}
  \end{center}
\end{figure}

  For some values of $\alpha$ and $p$, it is found that $\Psi =
\frac{1}{\sqrt{2}}(a_{+}+a_{-})$. So the `top' and `bottom' states co-exist
and, although each is fairly parallel on its own side of the torus, the two do
not mesh smoothly where they meet. Consequently, there are $2p$ defects each of
index $\frac{1}{n}$ around the torus's outer `rim' and $2p$ of index
$\frac{-1}{n}$ round the inside of the hole: a total of $4p$ defects. Hence as
$\alpha$ varies and $p$ jumps between integer values as described above,
vortex-antivortex quartets are formed or destroyed. Let these symmetrical
states be labelled $n^S_{p}$. The $4^S_{3}$ state for
$\alpha=\sqrt{2}$ is represented in Fig.(\ref{n4p3}).

  For other values of $\alpha$ and $p$, where the overlap between $\phi_{p}$
and $\phi_{-p}$ is large and hence the defects in the field become expensive,
the up-down symmetry is spontaneously broken, so that $\Psi=\phi_{p}$ or
$\Psi=\phi_{-p}$. Let these broken-symmetry states be labelled
$n^B_{p}$. As $\alpha$ varies, there are transitions between S and B states,
which, unlike the transitions betwen states of differing $p$ described in
section \ref{axisym}, may be regarded as true phase transitions within
mean-field theory, although they may be blurred by the inclusion of
fluctuations. As a rule of thumb, the $\psi^4$ term chooses S or B on the
basis of which covers the surface more uniformly.

  A table, giving the values of $\alpha$ for all transitions between
ground-state field configurations on the torus, is produced in the appendix.

  The free energy is next minimized with respect to $a$, giving the overall
magnitude of the order parameter field. Below the continuous transition, the
free energy is a minimum when
\begin{equation}
\label{size}
  a^2 = \frac{-\left(r+\frac{4\pi C \lambda_{0}}{\cal A}\right)}{uJ} .
\end{equation}

\subsection{Sphere}

  It was explained in section \ref{sphere} that the spherical Landau
levels are
degenerate with respect to the positions of the vortices. The
$\psi^4$ term lifts this degeneracy and makes the vortices repel. Hence,
following Park et al. \cite{Park92}, the ground state configuration is taken to
be that of highest symmetry. So, when $n=1$ the two zeros are antipodal. When
$n=2$, the zeros lie at the corners of a regular tetrahedron and, when $n=6$,
at the corners of an icosahedron.
The values of $J_{(\mbox{\tiny sph})}(n)$ were found using these
configurations. For $n=4$, the
defects lie at the corners of a cube whose top face has been rotated through
$45^o$ relative to the bottom face. The distance between these two faces is
then varied to minimize $J$. The optimal separation is found to be very close
to the length of a side of the original cube.

\section{The Phase Diagrams}
\label{PD}

  The minimized free energy of the sphere is now completely determined.
Expressing it as a fraction of the bending rigidity gives an equation relating
just four dimensionless combinations of thermodynamic parameters:
\begin{eqnarray}
\label{Fsph}
  \left[\frac{F_{(\mbox{\tiny sph})}}{T\,\kappa}\right] &=&
  4\pi\left(2+\left[\frac{\kappa_{G}}{\kappa}\right]\right)
  -2\pi\left[\frac{C^2}{{\cal A}u\kappa}\right]
  \frac{\left(\frac{1}{4\pi}\left[\frac{{\cal A}r}{C}\right]+n\right)^2}
  {J_{(\mbox{\tiny sph})}(n)}
  \;\;\;\;\; \mbox{for }\;\;\; r < \frac{-4\pi Cn}{\cal A} , \nonumber\\
  &=& 4\pi \left(2+\left[\frac{\kappa_{G}}{\kappa}\right]\right)
  \hspace{2.25in} \mbox{otherwise.}
\end{eqnarray}

\begin{figure}
  \epsfxsize=15cm
  \begin{center}
  \leavevmode\epsfbox{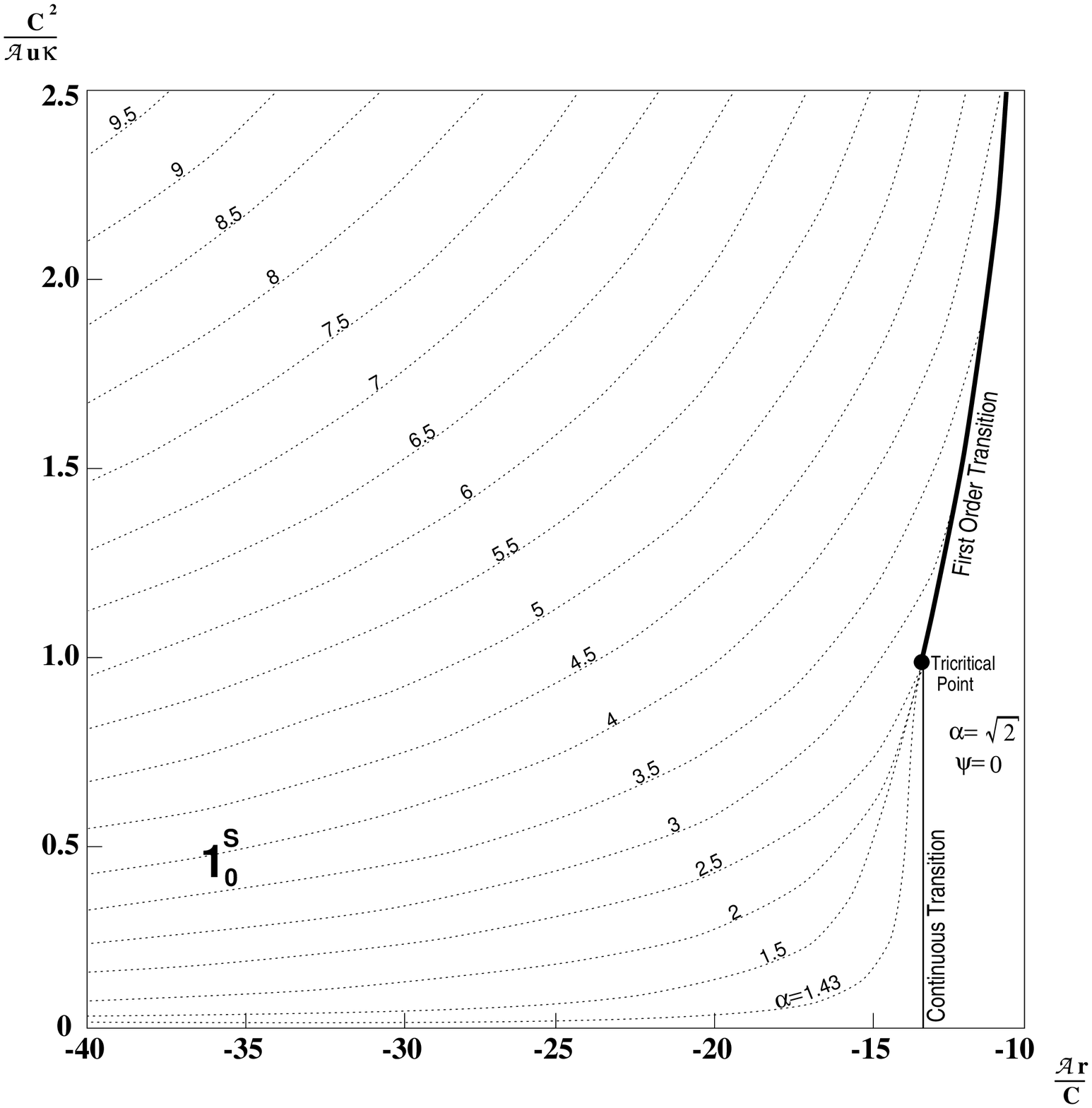}
  \caption{Phase diagram for a phase space restricted to axisymmetric, circular 
  cross-section tori of constant area ${\cal A}$ and variable aspect ratio
  $\alpha$. Vesicles in the low-temperature phase have intrinsic vector ($n=1$)
  order. See Eq.(\ref{F}) for definitions of the thermodynamic parameters.}
  \label{phasediags1}
  \end{center}
  \epsfxsize=15cm
  \begin{center}
  \leavevmode\epsfbox{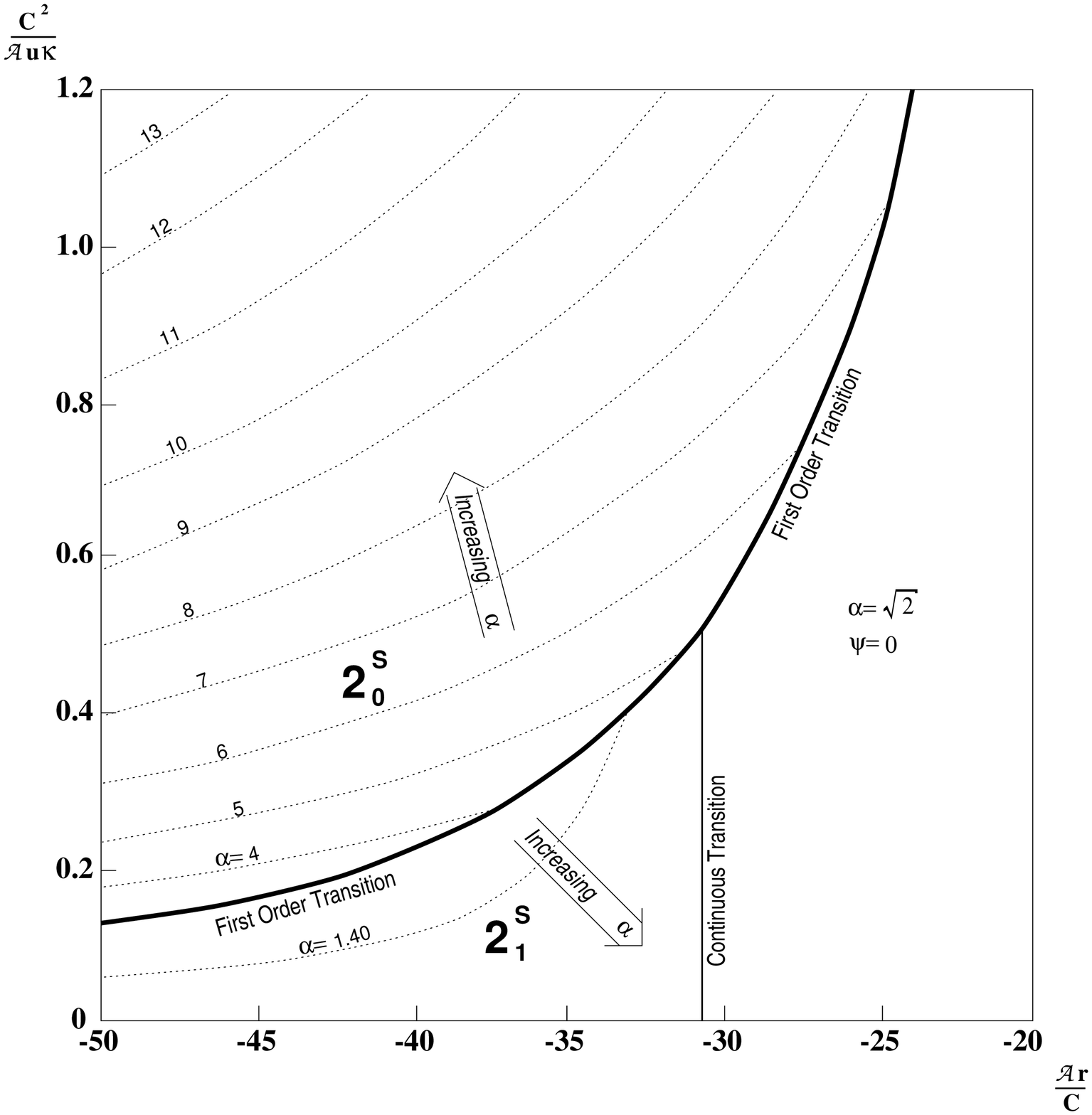}
  \caption{Phase diagram of axisymmetric, circular cross-section tori with
  intrinsic nematic ($n=2$) order in the low temperature phases.}
  \end{center}
  \epsfxsize=15cm
  \begin{center}
  \leavevmode\epsfbox{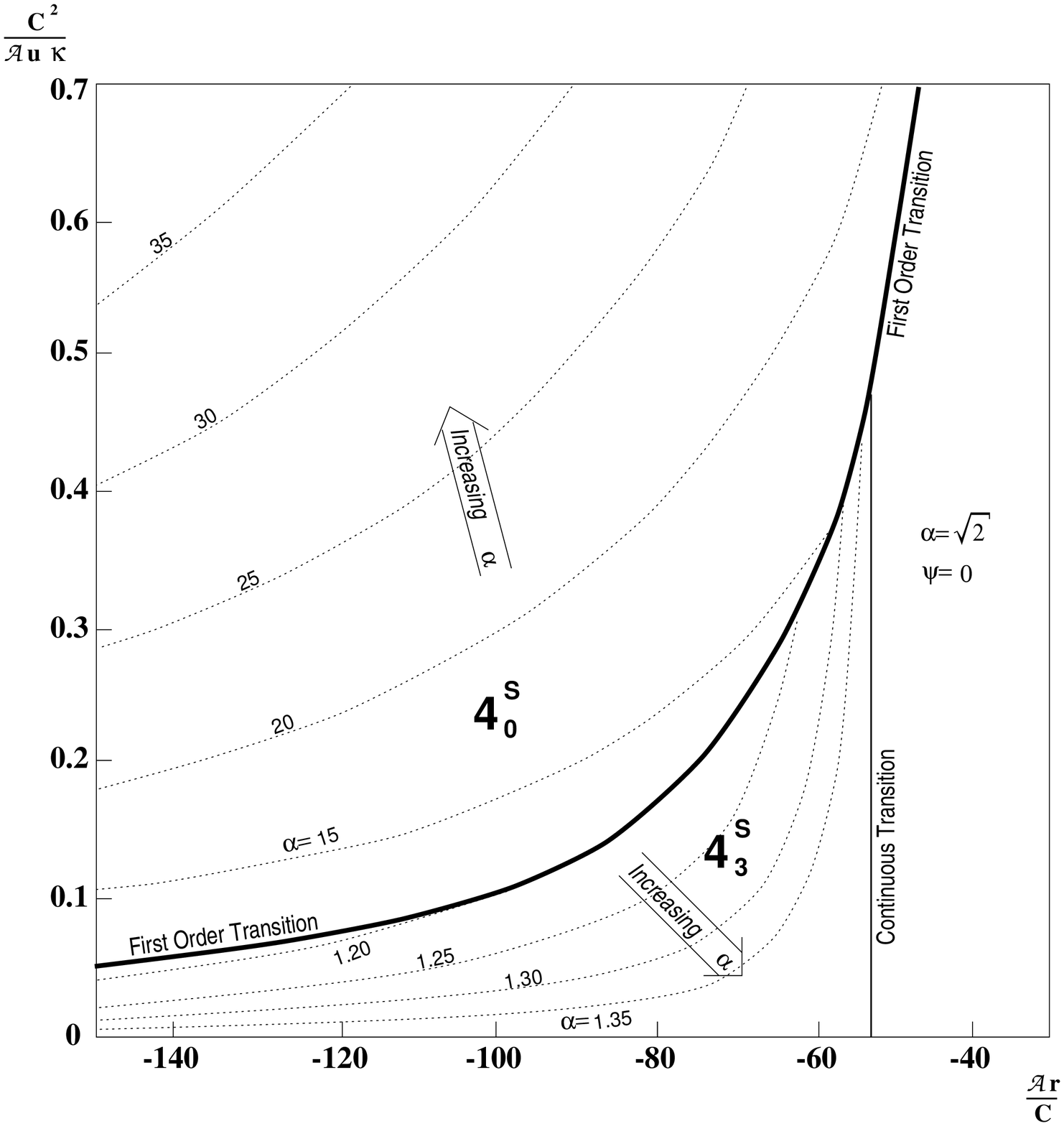}
  \caption{Phase diagram of axisymmetric, circular cross-section tori with
  intrinsic `tetratic' ($n=4$) order in the low temperature phases.}
  \end{center}
  \epsfxsize=15cm
  \begin{center}
  \leavevmode\epsfbox{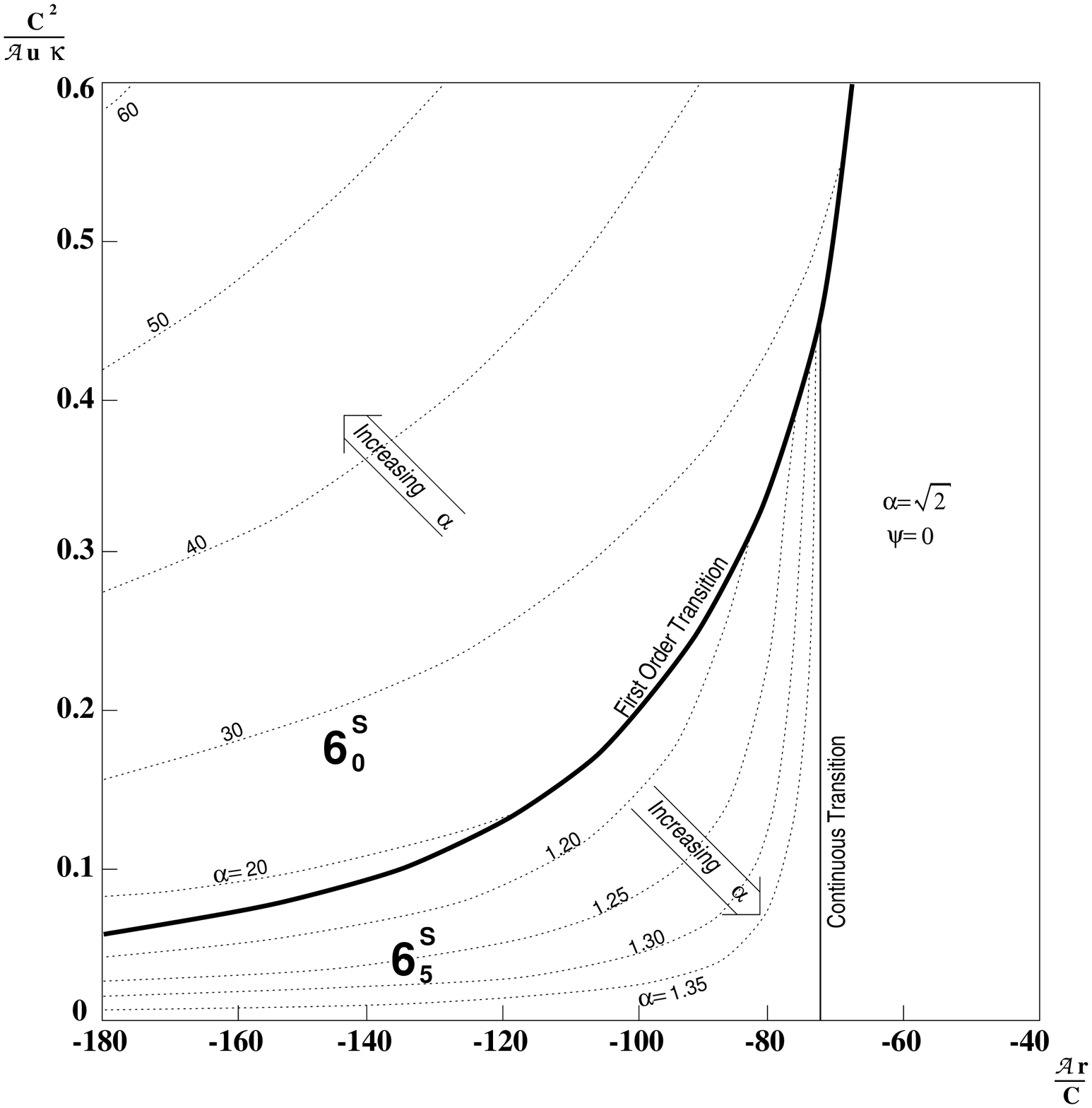}
  \caption{Phase diagram of axisymmetric, circular cross-section tori with
  intrinsic hexatic ($n=6$) order in the low temperature phases.}
  \label{phasediags1end}
  \end{center}
\end{figure}

  On the torus, the free energy remains a function of $\alpha$ only:
\begin{eqnarray}
\label{Ftor}
  \left[\frac{F_{(\mbox{\tiny tor})}}{T\,\kappa}\right] &=&
  \frac{2\pi^2\alpha^2}{\sqrt{\alpha^2-1}}
  -2\pi\left[\frac{C^2}{{\cal A}u\kappa}\right]
  \frac{\left(\frac{1}{4\pi}\left[\frac{{\cal A}r}{C}\right]
  +\lambda_{0}(\alpha)\right)^2}
  {J_{(\mbox{\tiny tor})}(\alpha,n)}
  \;\;\;\;\;\, \mbox{for }\;\;\; r < \frac{-4\pi C\lambda_{0}(\alpha)}
  {\cal A} , \nonumber\\
  &=& \frac{2\pi^2\alpha^2}{\sqrt{\alpha^2-1}}
  \hspace{2.575in} \mbox{otherwise}
\end{eqnarray}
which relates just three
dimensionless combinations of thermodynamic parameters. When Eq.(\ref{Ftor}) is
minimized with respect to $\alpha$, some of the physical richness of the
problem is lost, as many of the interesting phases and all the broken symmetry
states occur on tori of aspect ratios which are not the energetic optimum.
However, other interesting features arise, as will be seen in the phase
diagrams in Figs.(\ref{phasediags1})-(\ref{phasediags1end}), which
result from a phase space restricted
to only axisymmetric tori, and show contours of constant $\alpha$ as well as
lines of first order and continuous phase transitions. The phases are labelled
with the notation introduced earlier.
The diagrams are plotted
in the plane of the two independent combinations of thermodynamic parameters
in Eq.(\ref{Ftor}). In each case, to the far right of the diagram is the
high-temperature phase, for which $\psi=0$. In this phase, only the curvature
energy is non-vanishing so the aspect ratio is $\sqrt{2}$; the Clifford
torus. Note that the continuous transition, at which ordering arises, occurs
not at $r=0$, but is suppressed by the manifold's curvature to $r=r_{c}$, as
defined in section \ref{Landau} for the Clifford torus. Note that $r_{c}$
depends on aspect ratio, aspect
ratio depends on $\psi$ and $\psi$ depends on $r_{c}$. This non-linearity is
the cause of the first-order phase transitions. Consider for instance the $n=1$
diagram (Fig.(\ref{phasediags1})). A vesicle with small $\kappa$ is easily
deformed by the $1^S_{0}$-state ordering to
large aspect ratios at low temperatures. At large aspect ratios, $r_{c}$ is
close to zero. As the temperature rises, $\psi$ diminishes, and can no longer
hold the aspect ratio so far from the Clifford torus. As the aspect ratio falls,
ordering becomes more difficult. There comes a temperature at which the torus
jumps down a first order phase transition to the Clifford torus, where
$r_{c}< r$. At higher values of $\kappa$, the transition is continuous. The
$n=1$ phase diagram has a tri-critical point at $\frac{{\cal A}r}{C}=-13.68$
and $\frac{C^2}{{\cal A}u\kappa}\approx 0.98$. For higher values of $n$, the
first order transition persists for all values of $\kappa$, and meets the
continuous transition at a
multi-critical point. The continuous transition here is between the
high-temperature phase and the $n^S_{n-1}$ phase, for which
$\alpha<\sqrt{2}$ since the gradients in Fig.(\ref{lambdagraphs}) are
positive close to the Clifford torus.

\begin{figure}
  \epsfxsize=15cm
  \begin{center}
  \leavevmode\epsfbox{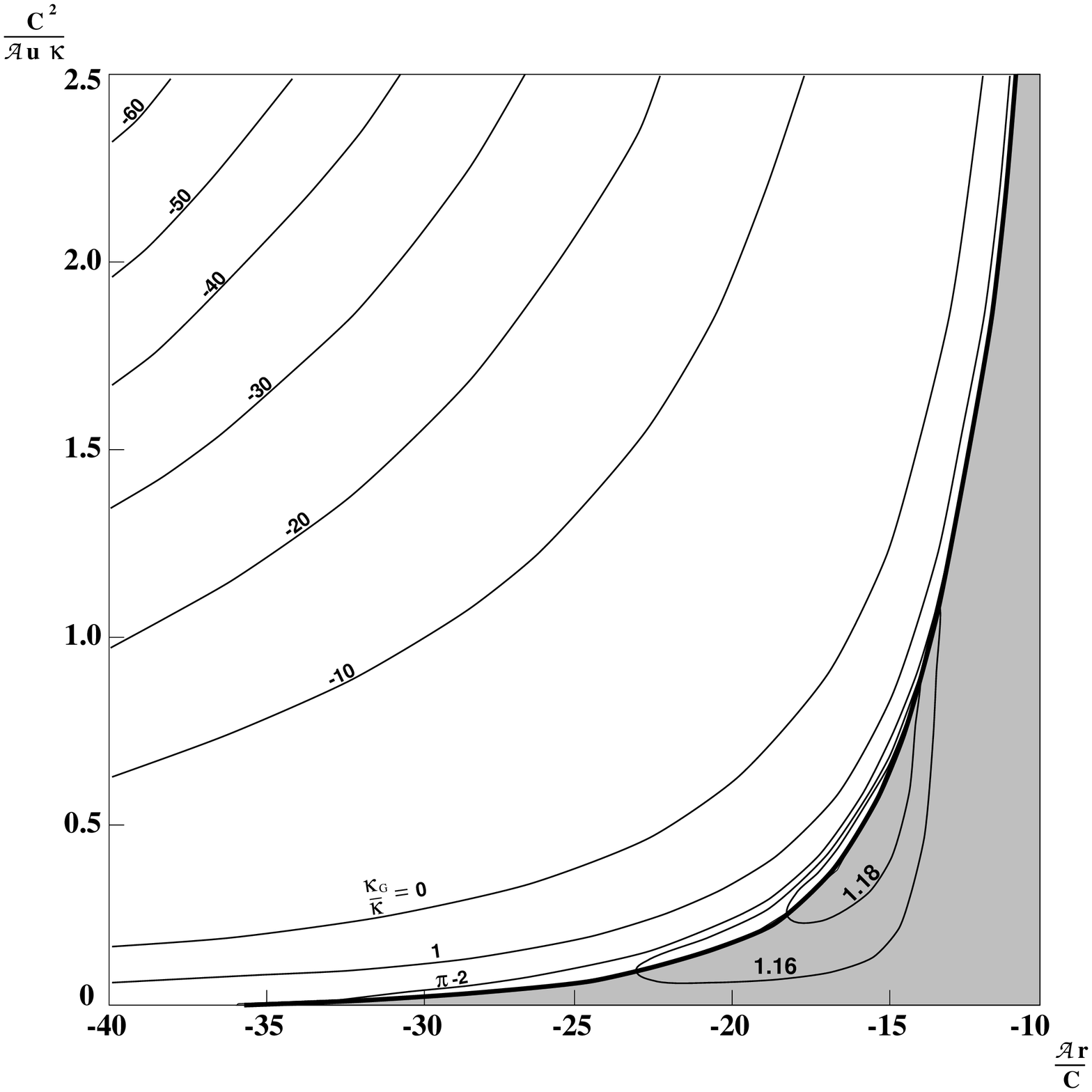}
  \caption{Phase diagram for constant area vesicles of genus zero or one, with
  intrinsic vector ($n=1$) order. The position of the transition line between
  genera depends on the ratio of intrinsic and total bending rigidities
  $\frac{\kappa_{G}}{\kappa}$. In the un-shaded region, genus one vesicles are
  axisymmetric, circular cross-section tori, whose aspect ratio is given in
  Fig(\ref{phasediags1}). Genus one vesicles that exist in
  the shaded region are non-axisymmetric. The area of the diagram occupied
  by spherical vesicles decreases with increasing $\frac{\kappa_{G}}{\kappa}$.}
  \label{phasediags2}
  \end{center}
  \epsfxsize=15cm
  \begin{center}
  \leavevmode\epsfbox{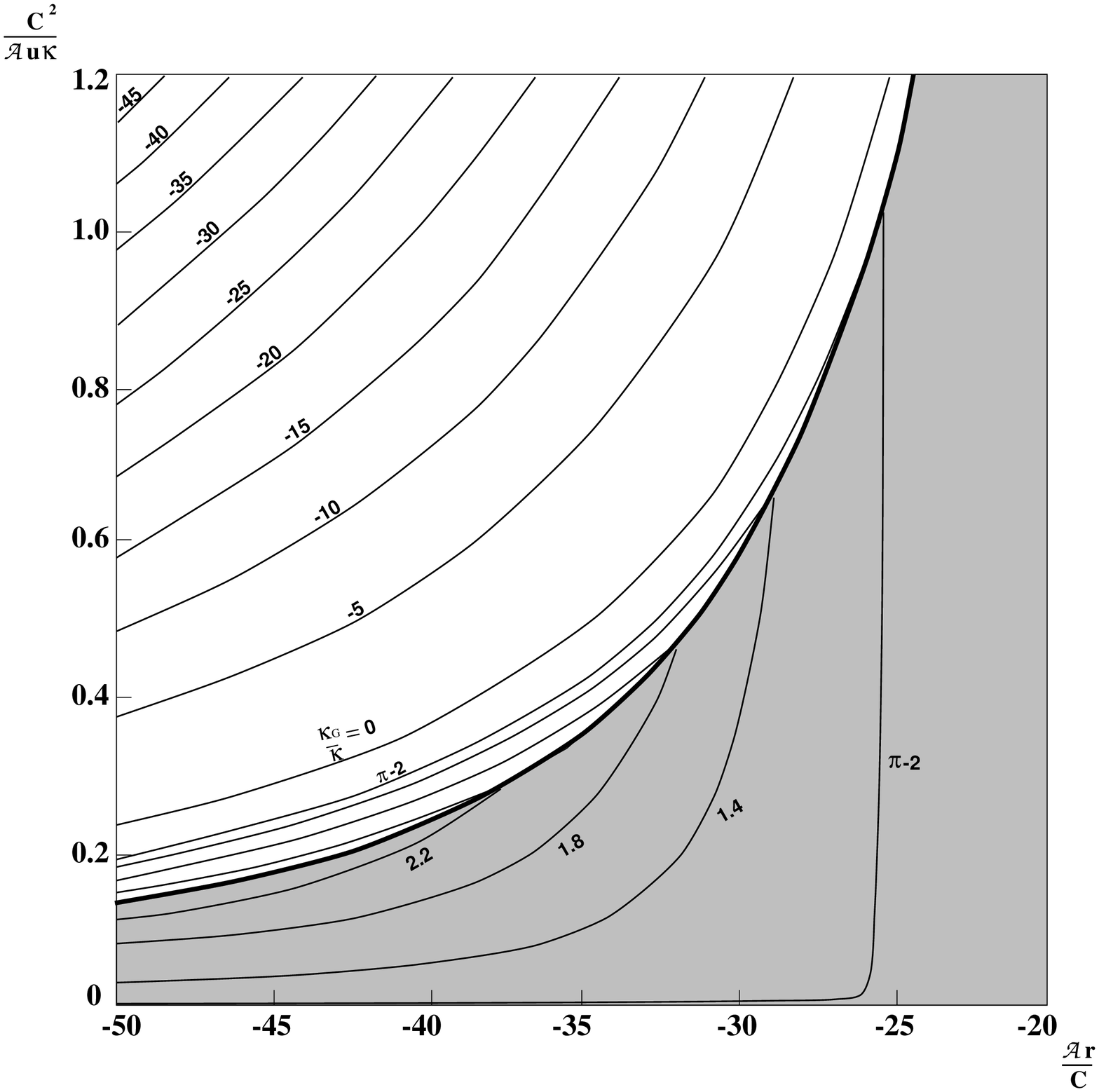}
  \caption{Phase diagram for constant area vesicles of genus zero or one, with
  intrinsic nematic ($n=2$) order.}
  \end{center}
  \epsfxsize=15cm
  \begin{center}
  \leavevmode\epsfbox{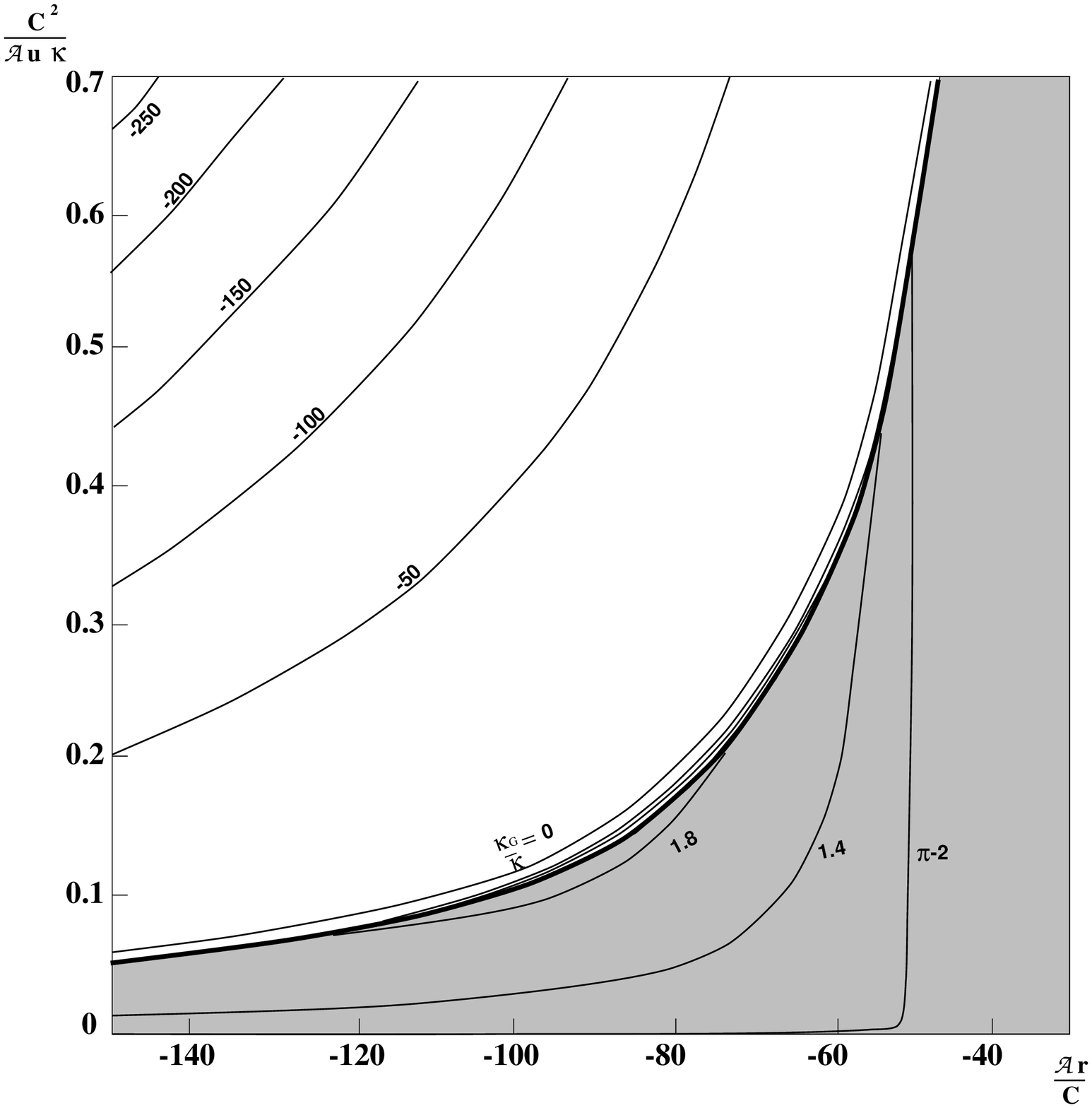}
  \caption{Phase diagram for constant area vesicles of genus zero or one, with
  intrinsic tetratic ($n=4$) order.}
  \end{center}
  \epsfxsize=15cm
  \begin{center}
  \leavevmode\epsfbox{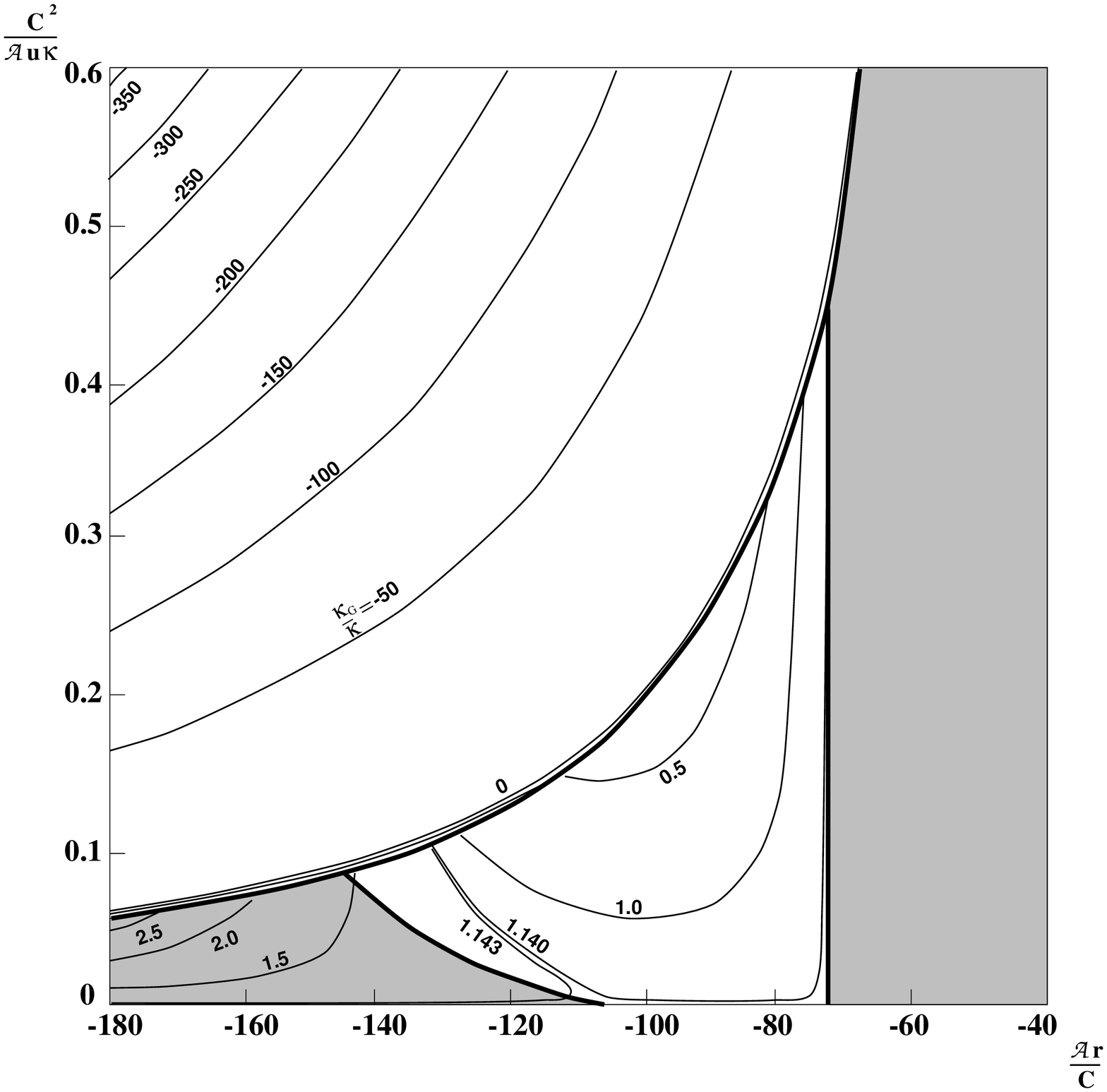}
  \caption{Phase diagram for constant area vesicles of genus zero or one, with
  intrinsic hexatic ($n=6$) order.}
  \label{phasediags2end}
  \end{center}
\end{figure}

  The free energy (Eq.(\ref{F})) has now been minimized with respect to the
order
parameter field's configuration and magnitude and with respect to the toroidal
aspect ratio. It remains only to minimize with respect to topology, according
to the prescription given in section \ref{non-axisym}.
Figs.(\ref{phasediags2})-(\ref{phasediags2end}) show the resulting phase
diagrams, which include
spherical, and axisymmetric and non-axisymmetric toroidal phases of vesicles.
No $\alpha$ contours are shown in these figures. The positions of the
transitions vary with $\frac{\kappa_{G}}{\kappa}$, and transition lines are
shown for various values of this quantity.
Figs.(\ref{phasediags2})-(\ref{phasediags2end})
should therefore each be regarded as several phase diagrams superimposed.
The $n^S_{n-1}$ phases do not persist when this larger phase space is explored,
except for $6^S_{5}$, for which $\lambda_{0}$ is less than the sphere value of
$6$, when $\alpha$ is close to $\sqrt{2}$. The diagrams are each divided into
shaded and unshaded regions by a heavy line. In the shaded regions, vesicles
are either spheres or non-axisymmetric tori while, in unshaded regions, either
spheres or axisymmetric tori occur.

  Consider first the $n=1$, $n=2$ and $n=4$ phase diagrams.
A transition line, whose
position depends on $\frac{\kappa_G}{\kappa}$ separates tori from spheres.
Note that for
$\frac{\kappa_G}{\kappa}>\pi-2$, the transition line is closed, with spheres
on the inside and tori of one sort or another on the outside, while for
$\frac{\kappa_G}{\kappa}<\pi-2$, the line is open, with spheres on the
high-temperature side and axisymmetric tori on the low-temperature side. This
magical value of $\pi-2$ is the critical value of $\frac{\kappa_G}{\kappa}$ at
which spheres would become tori if the model described curvature energy alone
and did not include terms in $\psi$. Note that in these three phase diagrams
the separatrix (the $\frac{\kappa_G}{\kappa}=\pi-2$ transition) has a vertical
limb at the value of $\frac{{\cal A}r}{C}$ at which ordering first forms on the
sphere, {\em ie.\ }at $r=r_{c}^{\mbox{\tiny sphere}}$.

  The $n=6$ diagram in Fig.(\ref{phasediags2end}) is a little more
complicated since
there are two distinct phases of ordering on axisymmetric tori. Again, the
diagram is divided into shaded and unshaded regions. The shaded region is now
split into two parts. The part on the right contains spheres if
$\frac{\kappa_G}{\kappa}<\pi-2$, and non-axisymmetric tori otherwise. The part
on the left contains spheres on the high $\frac{C^2}{{\cal A}u\kappa}$ side of
the transition line (whose position depends, of course, on
$\frac{\kappa_G}{\kappa}$) and non-axisymmetric tori on the other side. The
un-shaded region, which contains axisymmetric tori and spheres, is also divided
into two parts. The part in the top left of the diagram behaves as described
above for the other phase diagrams. The other part may (depending on
$\frac{\kappa_G}{\kappa}$) contain a closed transition line, which, unlike the
other closed transition lines we have met, has spheres on the outside and
axisymmetric tori, of small aspect ratio and $6_{5}^S$ field configuration, on
the inside.

  The four phase diagram
of Figs.(\ref{phasediags2})-(\ref{phasediags2end}) may
seem complicated but they
are easily understood if one remembers that: {\em a)} they are several diagrams
superimposed, and {\em b)} the regions in which spherical vesicles exist
become smaller as
$\kappa_{G}$ increases, since spheres have more intrinsic curvature than tori.
Having established from these diagrams that an axisymmetric torus exists in a
certain region, its aspect ratio may be found from
Figs.(\ref{phasediags1})-(\ref{phasediags1end}).

  It should be mentioned that the curves defining the edges of the shaded
regions in Figs.(\ref{phasediags2})-(\ref{phasediags2end}) have been
slightly smoothed by hand to remove
pixcellation which arose from numerical rounding errors.

\section{Validity of the Approximations}
\label{validity}

  Some order-of-magnitude calculations are now presented to give an indication
of the regimes of validity of the various approximations used.

\subsection{Lowest Landau Level Approximation}

  The Landau levels form a complete set of orthogonal functions in which to
expand the order parameter; $\psi=\sum_{p}a_{p}\phi_{p}$. Each eigenfunction
$\phi_{p}$ has an eigenvalue $\lambda_{p}$ defined by Eqs.(\ref{eigen}) and
(\ref{lambda}). The lowest Landau levels alone may be used if $a_{p}\ll a_{q}$
where $\lambda_{p}\neq\lambda_{q}=\lambda_{0}$. As an indicator of the region
in which this is valid, let us calculate $<a_{p}^{*}a_{p}>$ for Gaussian
fluctuations above the continuous transition, using the quadratic Hamiltonian
\begin{equation}
\label{quadratic}
  {\cal H} = \sum_{p}(r-\gamma\lambda_{p})b_{p}^{*}b_{p} + {\cal O}(b^{4})
\end{equation}
where $\gamma\equiv-\frac{4\pi C}{\cal A}$ and $b_{p}\equiv \sqrt{\frac{\cal A}
{4\pi}}a_{p}$.
The partition function is
\begin{displaymath}
  {\cal Z} = \int \exp\sum_{p}(\gamma\lambda_{p}-r)b_{p}^*b_{p}\:
  \prod_{p} db_{p}^*db_{p} .
\end{displaymath}
Hence
\begin{displaymath}
  <b_{p}^* b_{p}> = \frac{\pi}{r-\gamma\lambda_{p}}.
\end{displaymath}
So for the lowest landau levels, $<b_{p}^*b_{p}>_{0}=\frac{\pi}{r-r_{c}}$ and
for the next lowest level, $<b_{p}^*b_{p}>_{1}=\frac{\pi}{r-\gamma\lambda_{1}}$.
Hence the ratio of correlators is
\begin {displaymath}
  {\cal R} \equiv \frac{<a_{p}^*a_{p}>_{1}}{<a_{p}^*a_{p}>_{0}} =
  \frac{<b_{p}^*b_{p}>_{1}}{<b_{p}^*b_{p}>_{0}} = \frac{x}{x+4\pi(\lambda_{1}
  -\lambda_{0})}
\end{displaymath}
where $x=\frac{{\cal A}(r-r_{c})}{C}$ is the x-coordinate in the phase
diagrams, measured with respect to the transition. We see that ${\cal R}$ is
small, and hence the lowest Landau level approximation is good, when
$|x|\tilde{<}4\pi(\lambda_{1}-\lambda_{0})$, the right hand side of which may be
estimated from Fig.(\ref{lambdagraphs}). For the Clifford torus,
$(\lambda_{1}-\lambda_{0})\sim 0.5$ so the approximation is quantitatively
accurate only within $\sim$3 of the transition, along the x-axis. However,
Figs.(\ref{phasediags1})-(\ref{phasediags2end}) cover a much larger region than
this, in order to clarify the qualitative features of the diagrams, for which
the calculation at this level of approximation is a useful guide. It should be
noted that, at `transitions' from one lowest Landau level configuration to
another, $(\lambda_{1}-\lambda_{0}) \rightarrow 0$, so the approximation must
break down here, as noted in section \ref{axisym}. Also, in the large-$\alpha$
phases below the first-order transitions, $r_{c}$ is higher, so the system is
far from the lowest Landau level regime. These phases are included in the
diagrams in order to show the first-order transitions.

\subsection{Mean Field Theory}

  For this order-of-magnitude analysis, let us treat two parts of the free
energy separately: first we will demand that fluctuations in the Helfrich
Hamiltonian may be neglected in the absence of orientational order, then,
within this fixed, non-fluctuating environment, it will be required that the
terms in $\psi$ satisfy the Ginzburg criterion for the validity of
mean-field theory.

  In a simplistic calculation, good over only a short range, the Monge gauge is
used to parameterize a locally horizontal surface, defining its
z-coordinate as a function of x and y. Keeping terms of second order in z, and
expanding z in plane-wave modes: $z\sim{\cal A}^{\frac{1}{2}}\sum_{k}c_{k}
e^{ik\cdot x}$ gives
\begin{displaymath}
  \kappa\int (K_{a}^a)^2 d{\cal A} \approx \kappa\int (\nabla^2 z)^2 dx dy
  = \kappa {\cal A}^2 \sum_{k} |c_{k}|^2 k^4 .
\end{displaymath}
Note that the Gaussian curvature energy does not influence fluctuations as it
is a topological invariant. Hence
\begin{displaymath}
  <|c_{k}|^2> = \frac{1}{\kappa {\cal A}^2 k^4}
\end{displaymath}
so, for fluctuations to be negligible on the scale of the system size requires
$\kappa\gg\frac{1}{4\pi^2}$ .  This is intuitively obvious; that a large
bending rigidity is required to combat fluctuations. In fact, this simple
calculation on a flatish surface is over-stringent, since intrinsic curvature,
as exists in closed surfaces, reduces the size of fluctuations \cite{?}.

  The Ginzburg criterion is now computed for the Ginzburg-Landau part of the
free energy. This is a comparison between the size of the discontinuity in
specific heat at the phase transition, calculated with mean-field theory, and
the size of the divergence in specific heat due to Gaussian fluctuations. Close
to the transition, $r$ goes approximately linearly with temperature $T$. Hence
$(r-r_{c})=\rho(T-T_{c})$, where $T_{c}$ is the shifted mean-field transition
temperature, and $\rho$ is some constant. From Eq.(\ref{Fsph}) it follows that
the mean-field discontinuity in specific heat is
\begin{displaymath}
  \Delta C = \frac{\rho^2 T_{c}^2}{4\pi u J} .
\end{displaymath}
For simplicity, the specific heat of fluctuations is calculated for the sphere,
above the phase transition. Using the quadratic form Eq.(\ref{quadratic}), and
only the $(2n+1)$ lowest Landau levels, gives
a free energy density $f=\frac{(2n+1)T}{\cal A}(\ln \rho(T-T_{c})-\ln\pi)$.
Hence the singular part or the specific heat is
\begin{displaymath}
  C_{\mbox{\tiny sing.}}=\frac{(2n+1)\rho}{{\cal A}(1-\frac{T}{T_{c}})^2} ,
\end{displaymath}
giving a critical exponent $\alpha=2$ since we are in a zero dimensional
$k$-space. The ratio of these two specific heats is
\begin{displaymath}
  \frac{C_{\mbox{\tiny sing.}}}{\Delta C} =
  \left(\frac{\zeta}{1-\frac{T}{T_{c}}}\right)^2
\end{displaymath}
where
\begin{displaymath}
  \zeta = \sqrt{\frac{4\pi(2n+1)uJ}{{\cal A}\rho T_{c}}}
\end{displaymath}
and mean field theory is good for $|r-r_{c}|\tilde{>}\zeta$ . (N.B.\ This
condition is equivalent to demanding that tree graphs dominate over one-loop
graphs.) So the threshold,
in the phase diagrams, between the mean-field regime and the
fluctuation-dominated regime is given by the curve $x^2 y=(2n+1)/\kappa$,
where $x\equiv \frac{{\cal A}(r-r_{c})}{C}$ and $y\equiv \frac{C^2}{{\cal A}u
\kappa}$, having set $\rho\sim T_{c}^{-1}$ and $J\sim\frac{1}{4\pi}$. At a
height $y_{0}$, this region has width
$\Delta=\sqrt{\frac{4\pi(2n+1)J}{\kappa y_{0}}}$. This should be much smaller
than the region of interest {\em ie.\ }that in which the lowest Landau level
approximation is valid. So $\Delta \ll 3$ and, by inspection, in all the
phase diagrams, we are interested in $y_{0}\sim \frac{1}{n}$. Consequently the
condition for a mean-field treatment of the order parameter to be valid in this
investigation is just $\kappa\gg\frac{n^2}{9}$, which is always consistent with
the validity of a mean-field treatment of the membrane shape, calculated above.

\subsection{The Ginzburg-Landau Model}

  The Ginzburg-Landau free energy functional is and expansion in $\psi$,
truncated at order $\psi^4$. This is valid when ${\cal A}|\psi|^2\ll 1$. In
the mean-field regime, this is always the case above the continuous transition.
Below the transition, $\psi\sim\frac{|r-r_{c}|}{uJ\sqrt{4\pi}}$, from
Eq.(\ref{size}). So the inequality is marginal when $y\sim\frac{C}{\pi{\cal A}
\kappa x}$ where $x$ and $y$ are given above. This defines a region in the
phase diagrams of width $\Delta=\frac{Cn}{\pi{\cal A}\kappa}$ at a height
$y=\frac{1}{n}$, which is in the region of interest. Again, let's demand that
$\Delta>>3$ to be consistend with the region of validity of the lowest Landau
level approximation. Consequently, it is required that $\frac{C}{\cal A} \gg
\frac{3\pi}{n}\kappa$.

  Dropping constants of order unity, we find that mean-field theory, the
Ginzburg-Landau model and the lowest Landau level approximation are all valid
in a region of half-width $\sim 3$ in the phase diagrams, given that
\begin{displaymath}
  \frac{C}{\cal A} \gg \kappa \gg 1 .
\end{displaymath}

\subsection{The Shapes}

  Close to the ordering transition, $\psi$ is very small, so the order cannot
deform the torus far from
Clifford's aspect ratio $\sqrt{2}$, or the genus-zero shape far from a sphere,
so the vesicle shapes used in this theoretical investigation are undoubtedly
good. As Park et al.\ discovered in \cite{Park92}, $\psi$ wants to live in a
flat space, and can achieve a flattening of the membrane via the gauge-like
coupling in Eq.(\ref{F}). This it will do where it is large, at the expense of
an increased curvature where $\psi$ is small {\em eg.\ }at the defects. This
pay-off is unavoidable, since the integral of intrinsic curvature is
topologically fixed. Hence spherical vesicles are slightly deformed by the
in-plane order, into rounded polyhedra, whose vertices are at the order
parameter's vortices. This is the effect which has been ignored in this paper.
The same process will also occur in tori, especially in those parts of the
phase diagrams where ordering is predicted to deform the tori to large aspect
ratios. In general, one might expect that the larger the aspect ratio, the
worse the approximations of circular cross-section and axisymmetry. But nature
is on our side here. Since, in these large-aspect-ratio tori, $\psi$ is
always in the $n_{0}^S$ configuration, which has no vortices, and becomes very
uniform for large aspect ratios. So, although this approximation results in
some quantitative inaccuracies, they are likely to be small. Also, they must
change the free energy of the torus and the sphere, if not by the same amount,
then at least in the same direction. So the phase diagrams may not have exactly
the right numbers, but they do have more-or-less the right shape.

\section{Conclusion}

  It is clear that orientational order in a toroidal geometry is a subject rich
in physical diversity, and deserving of further study in condensed matter
theory. Also, introducing orientational order into the surface is an
interesting way of stabilizing a torus against the massless Goldstone mode of
conformal transformations, without resorting to spontaneous curvature or a
constraint of constant volume or bilayer area difference. As far as experimental
verification is concerned, the principal results of this paper are embodied in
the final four phase diagrams. The approximations used in their derivation may
result in quantitative inaccuracies, but the main qualitative features of these
diagrams are predicted to be seen in experiment in the future. At present an
experimental realization is somewhat difficult, due to the fragile
nature of laboratory-produced vesicles.

\section{Acknowledgements}

  I would like to thank Prof.\ Mike Moore for his guidance in this work. I
am grateful for the helpful comments of Tom Lubensky, Udo Seifert, Renko de
Vries and the numerous others with whom I have conversed on the
subject. I wish to acknowledge the EPSRC for my funding.

\appendix
\section{Table of Mean-Field Phase Transitions}

The axisymmetric toroidal aspect ratios at which the order parameter
field, which minimizes the free energy in the regime of the lowest Landau level
approximation, changes its configuration between states of different $p$ or
different symmetry are given in the table below. See sections \ref{axisym}
and \ref{axisym2} for explanations of these transitions.
Note that an $n=1$ (vector) field has no such transitions, as its ground state
is always $1_{0}^S$.
\vspace{1cm}
\begin{center}
\begin{tabular}{||c|c||}	\hline \hline
{\bf Transition}		& {\bf Aspect Ratio}   \\ \hline \hline
$2_{1}^S \rightarrow 2_{1}^B$	& 2.31 \\
$2_{1}^B \rightarrow 2_{0}^S$	& 2.41 \\ \hline
$4_{3}^S \rightarrow 4_{2}^S$	& 3.34 \\
$4_{2}^S \rightarrow 4_{1}^B$	& 4.56 \\
$4_{1}^B \rightarrow 4_{0}^S$	& 4.98 \\ \hline
$6_{5}^S \rightarrow 6_{4}^S$	& 3.15 \\
$6_{4}^S \rightarrow 6_{3}^S$	& 5.46 \\
$6_{3}^S \rightarrow 6_{2}^B$	& 6.63 \\
$6_{2}^B \rightarrow 6_{1}^B$	& 7.24 \\
$6_{1}^B \rightarrow 6_{0}^S$	& 7.83 \\ \hline \hline
\end{tabular}
\end{center}

\end{document}